\documentclass[a4paper,11pt]{article}
\makeatletter
\newcommand\notsotiny{\@setfontsize\notsotiny{6.8}{7.8}}
\makeatother

\usepackage[DIV12]{typearea}
\usepackage{booktabs}
\usepackage{amsmath}
\usepackage{amssymb}
\usepackage{bbm}
\usepackage[utf8]{inputenc}
\usepackage{xspace}
\usepackage[caption=false]{subfig}
\usepackage{nicefrac}
\usepackage{graphicx}
\usepackage{cite}  
\usepackage{nccfoots}
\usepackage[graphicx]{realboxes}
\usepackage{float}

\usepackage{multirow}
\usepackage[hang, small,labelfont=bf,up,textfont=it,up]{caption}

\newcommand{\rep}[1]{\ensuremath\boldsymbol{#1}}
\newcommand{\crep}[1]{\ensuremath\bar{\boldsymbol{#1}}}
\newcommand{\Z}[1]{\ensuremath{\mathbbm{Z}_{#1}}} % z_N ->\Z{N}
\newcommand{\SO}[1]{\ensuremath{\mathrm{SO}(#1)}}
\newcommand{\SU}[1]{\ensuremath{\mathrm{SU}(#1)}}

\newcommand{\U}[1]{\ensuremath{\mathrm{U}(#1)}}
\newcommand{\E}[1]{\ensuremath{\mathrm{E}_{#1}}}
\newcommand{\I}{\mathrm{i}}
\newcommand{\Id}{\mathbbm{1}}

\newcommand{\nphantom}[1]{\sbox0{#1}\hspace{-\the\wd0}}

\usepackage{xcolor}

\definecolor{darkgreen}{HTML}{109930}

\advance \headheight by 3.0truept       % for 12pt mandatory...
\usepackage[pdftex]{hyperref}

\hypersetup{
    pdftitle = {Predicting the orbifold origin of the MSSM},
    pdfauthor = {Parr, Vaudrevange, Wimmer}
}
 
\begin{document}

\begin{titlepage}

\begin{flushright}
\normalsize{TUM-HEP 1253/20}
\end{flushright}

\vspace*{1.0cm}

\begin{center}
{\Large\textbf{Predicting the orbifold origin of the MSSM}}

\vspace{1cm}

\textbf{Erik~Parr}, \textbf{Patrick~K.S.~Vaudrevange} and \textbf{Martin~Wimmer}
\Footnote{*}{%
\href{mailto:erik.parr@tum.de;patrick.vaudrevange@tum.de;martin.wimmer@tum.de}{\tt erik.parr@tum.de, patrick.vaudrevange@tum.de, martin.wimmer@tum.de} 
}
\\[5mm]
Physik Department T75, Technische Universit\"at M\"unchen,\\
James-Franck-Stra\ss e 1, 85748 Garching, Germany\\
\end{center}

\vspace{1cm}

\begin{abstract}
MSSM-like string models from the compactification of the heterotic string on toroidal orbifolds 
(of the kind $\mathbbm{T}^6/P$) have distinct phenomenological properties, like the spectrum of 
vector-like exotics, the scale of supersymmetry breaking, and the existence of non-Abelian flavor 
symmetries. We show that these characteristics depend crucially on the choice of the underlying 
orbifold point group $P$. In detail, we use boosted decision trees to predict $P$ from 
phenomenological properties of MSSM-like orbifold models. As this works astonishingly well, we can 
utilize machine learning to predict the orbifold origin of the MSSM.
\end{abstract}

\end{titlepage}

\newpage

%%%%%%%%%%%%%%%%%%%%%%%%%%%%%%%%%%%%%%%%%%%%%%%%%%%%%%%%%%%%%%%%%%%%%%%%%%%%%%%%%%%%%%%%%%%%%%%%%%%%%%%%%%%%%%%%%%%%%%%%%%%%%%%%%%%%%%%%%
\section{Introduction}

String theory compactified to four-dimensional space-time naturally provides a unified framework for 
quantum gravity and gauge interactions with chiral matter. This fact raises the obvious question 
whether string theory can incorporate the Standard Model (SM) of particle physics (or its Minimal 
Supersymmetric Extension, the MSSM). A definite answer to this question would be given by an explicit 
construction of a string compactification that is in agreement with all experimental facts from 
particle physics (and, if one is even more ambitious, with all cosmological observations). However, 
due to the enormous number of four-dimensional string models~\cite{Lerche:1986cx,Douglas:2003um} 
and the computational complexity~\cite{Halverson:2018cio} a naive search in the string landscape 
for the MSSM is very likely to fail. New methods seem to be unavoidable to narrow down the string 
landscape towards realistic particle physics. 

In recent years, big data and machine learning (ML) has entered the field of 
strings~\cite{He:2017aed,Krefl:2017yox,Ruehle:2017mzq,He:2018jtw,RUEHLE20201}. Various tasks have 
been addressed, for example, to identify the topological structure of the string landscape using 
persistent homology~\cite{Cole:2018emh}, to predict the Hodge numbers of complete intersection 
Calabi-Yau manifolds (CICYs) with large $h^{1,1}$~\cite{Bull:2019cij}, to find consistent type 
IIA D6-brane configurations that yield MSSM-like models using Deep Reinforcement 
Learning~\cite{Halverson:2019tkf}, to identify whether a given CICY is elliptically fibered or 
not~\cite{He:2019vsj}, to explore the landscape of type IIB flux vacua using genetic 
algorithms~\cite{Cole:2019enn}, to find numerical metrics of Calabi-Yau manifolds by combining 
conventional curve fitting and techniques from supervised learning~\cite{Ashmore:2019wzb}, and to 
approximate K{\"a}hler metrics for type IIB Calabi-Yau compactifications using generative 
adversarial networks (GANs)~\cite{Halverson:2020opj}. Hence, encouraged by these results, 
techniques from big data and ML are expected to yield new insights into the string 
landscape.

In the context of heterotic orbifolds~\cite{Dixon:1985jw,Dixon:1986jc,Ibanez:1986tp}, the 
Mini-Landscape of $\Z{6}$-II orbifold models has been a test ground for model searches: First, 
individual models have been identified and analyzed~\cite{Kobayashi:2004ya,Buchmuller:2005jr,Buchmuller:2006ik,Lebedev:2007hv}. 
Then, larger scans have been performed~\cite{Lebedev:2006kn,Lebedev:2008un,Nilles:2014owa,Olguin-Trejo:2018wpw}. 
Finally, new methods from ML have been applied to this test ground, like an autoencoder 
neural network to automatically identify fertile islands in the $\Z{6}$-II Mini-Landscape~\cite{Mutter:2018sra} 
and techniques from contrast data mining to reduce the landscape by extracting new features of 
orbifold models that correlate with their phenomenological property of being MSSM-like~\cite{Parr:2019bta}. 
However, there are in total 138 Abelian orbifolds with $\mathcal{N}=1$ supersymmetry~\cite{Fischer:2012qj}. 
So, the heterotic orbifold landscape is much wider than $\Z{6}$-II. Consequently, how do we know 
that we will find the most promising models in the $\Z{6}$-II region of the landscape? Actually, 
can ML algorithms predict the orbifold geometry which most likely reproduces a certain MSSM-like 
model? As shown in this paper using a boosted decision tree, the answer to this question seems to 
be positive.

This paper is organized as follows: Section~\ref{sec:OrbifoldLandscape} begins with a brief 
review of the Orbifold-Landscape of all known MSSM-like orbifold models constructed so far. In 
addition, motivated by some generic properties of these models, we define phenomenological features 
that characterize MSSM-like models in general. Then, in section~\ref{sec:Boosteddecisiontree} 
we discuss boosted decision trees and present the resulting predictions for the orbifold origin of 
the MSSM in section~\ref{sec:results}. Finally, section~\ref{sec:conclusion} gives conclusions and 
outlook.
\enlargethispage{0.1cm}

\section{Phenomenology of the Orbifold-Landscape}
\label{sec:OrbifoldLandscape}

\begin{figure}[t]
\centering
\includegraphics[scale=1.4]{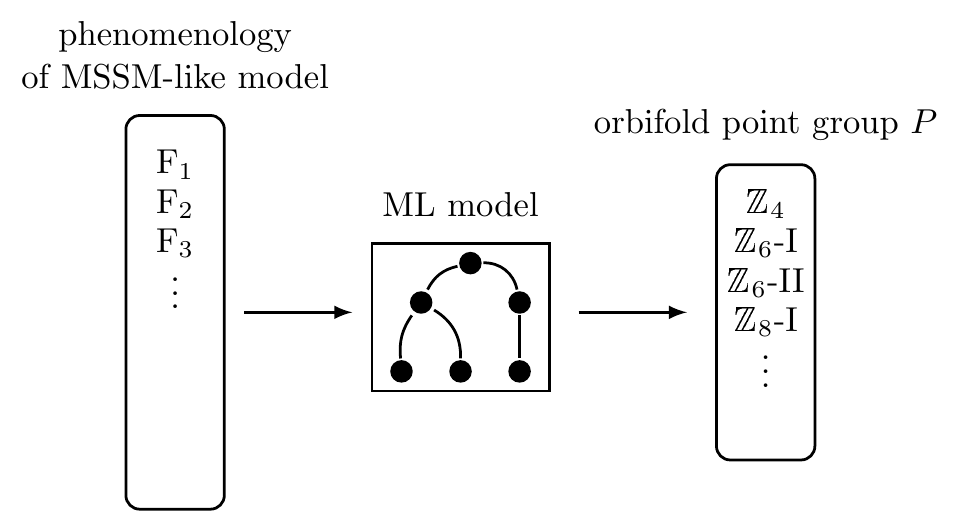}
\caption{We use a machine learning (ML) model to predict the orbifold point group $P$ that most 
likely is able to reproduce in string theory a given MSSM-like bottom-up model, which we specify 
by some phenomenological features F$_i$.
\label{fig:ml_overview}
}
\end{figure}

The aim of this paper is to construct a machine learning (ML) model that predicts the orbifold
origin of MSSM-like bottom-up models. In more detail, motivated by the generic features of 
MSSM-like string models in the Orbifold-Landscape, we train an ML model to predict the orbifold 
point group that has the highest probability to reproduce a given MSSM-like model, see 
figure~\ref{fig:ml_overview}. To train such an ML model, we need a large dataset of MSSM-like 
orbifold models based on various different orbifold point groups. Then, we have to define and compute 
some phenomenological features that yield a basic characterization of MSSM-like models. These 
features are taken by the ML algorithm as input, while the output of the ML algorithm 
is the prediction of the corresponding orbifold point group. Thus, in the following we first discuss 
our dataset of MSSM-like orbifold models and, afterwards, we define our phenomenological features.

\subsection{The dataset of MSSM-like orbifold models}

For the analysis we use a large dataset of 126,783 inequivalent MSSM-like string models originating 
from the $\E{8}\times\E{8}$ heterotic string, compactified on various orbifolds $\mathbbm{O}$ with 
Abelian point group $P$ (e.g.\ in the case without roto-translations, $\mathbbm{O}=\mathbbm{T}^6/P$, 
where $P$ is either $\Z{N}$ or $\Z{N}\times\Z{M}$, see ref.~\cite{Fischer:2012qj}). This dataset 
is obtained as follows: First, we take the MSSM-like orbifold models from the searches performed in 
refs.~\cite{Nilles:2014owa,Olguin-Trejo:2018wpw,Parr:2019bta}. Then, we construct new 
$\Z{N}\times\Z{M}$ MSSM-like models using the enhanced search algorithm based on contrast 
patterns~\cite{Parr:2019bta}. Finally, these individual datasets are merged using the 
orbifolder~\cite{Nilles:2011aj} resulting in 126,783 inequivalent MSSM-like orbifold 
models~\footnote{The MSSM-like orbifold models can be found as arXiv ancillary files, see 
refs.~\cite{Parr:2019bta} and~\cite{Parr:2020anc}.}. In the first row of table~\ref{tab:MSSMsummary1} 
we list the orbifold point groups $P$ that yield MSSM-like string models and the second row gives 
the number of inequivalent models based on the respective point group. Note that from the point of 
view of data analysis, our dataset is highly imbalanced: Some point groups (like $\Z{4}$) give rise 
to only a few hundred MSSM-like orbifold models, while others (like $\Z{2}\times\Z{4}$) yield several 
ten-thousand MSSM-like models. Moreover, since we only have one MSSM-like orbifold model based 
on the $\Z{7}$ point group, see refs.~\cite{RamosSanchez:2008tn,Olguin-Trejo:2018wpw}, we decide to 
leave out the $\Z{7}$ point group from our prediction task.

In the following, we analyze our dataset of MSSM-like orbifold models for their generic 
phenomenological properties and, by doing so, we identify some universal features of MSSM-like 
models.

\subsection{SM singlets}

MSSM-like orbifold models typically yield $\mathcal{O}(100)$, i.e.\ on average 146, SM singlets 
$s^0$ with SM charges $(\rep{1}, \rep{1})_0$ (see the eighth row in tables~\ref{tab:MSSMsummary1} 
and~\ref{tab:MSSMsummary2}). Generically, they are charged under several hidden $\U{1}$ factors and 
sometimes even under a non-Abelian hidden sector gauge group $G_\mathrm{hidden}$. The existence of 
these SM singlets has several implications: As we will discuss in detail in 
sections~\ref{sec:vectorlikeexotics} and~\ref{sec:AnomalousU1}, they can acquire non-vanishing 
vacuum expectation values (VEVs) $\langle s^0\rangle\neq 0$ without breaking supersymmetry, i.e.\ 
$F=D=0$. Moreover, interpreting these SM singlets as right-handed neutrinos, they can give rise 
to a seesaw mechanism with a model-dependent seesaw scale that is typically somewhat below the 
string scale~\cite{Buchmuller:2007zd}. Since, their presence is so generic in MSSM-like string 
models from heterotic orbifolds, we include their number into our feature space.

\subsection{Vector-like exotics}
\label{sec:vectorlikeexotics}

Beside three (chiral) generations of quarks and leptons, a Higgs-pair and SM singlets, MSSM-like 
orbifold models are generically equipped with vector-like exotics (i.e.\ matter that is charged 
under the SM and has a mass-partner transforming in the complex conjugate representation with 
respect to the SM). All types of vector-like exotics that appear in the Orbifold-Landscape are 
listed in tables~\ref{tab:MSSMsummary1} and~\ref{tab:MSSMsummary2} in appendix~\ref{app:exotics}. 
Since they appear so frequently, we choose the numbers of vector-like exotics of all kinds as 
features (except for the vector-like exotics that only appear in the unique MSSM-like $\Z{7}$ 
orbifold model). 

The MSSM itself is defined without any vector-like exotics beside the Higgs-pair. So, what are the 
phenomenological consequences if vector-like exotics are present? Let us denote a pair of 
vector-like exotics by $X$ and $\bar{X}$ in the following. In many cases, these exotics can become 
very massive through terms in the superpotential of the form $\mathcal{W} \supset (M_{\mathrm{Planck}})^{1-p}\,(s^0)^p\,X\,\bar{X}$. 
Here, $p \in\mathbbm{N}_+$ and $s^0$ denotes an MSSM singlet $(\rep{1}, \rep{1})_0$ that can acquire 
a large non-vanishing VEV. This would render the vector-like exotics $X$ and $\bar{X}$ massive, with 
a mass that can be close to the Planck scale $M_{\mathrm{Planck}}$ depending on the size of 
$\langle s^0\rangle$ and $p \in\mathbbm{N}_+$. Still, one can argue that the presence of vector-like 
exotics might be a virtue or a problem: On the one hand, new elementary particles with spin 0 or 1 
(called leptoquarks) have been proposed, especially to address some flavor anomalies, see for 
example refs.~\cite{Buchmuller:1986zs,Bauer:2015knc,Diaz:2017lit,Tanabashi:2018oca} and references 
therein. In this scenario, the leptoquark has to be light compared to the Planck scale. 
On the other hand, the presence of vector-like exotics can affect gauge coupling unification and 
it can yield severe cosmological problems. Thus, we decide to look for ``(almost) perfect 
MSSM-like models'' that have no (or a minimal amount of) vector-like exotics. In summary, our basic 
features correspond to the numbers of all types of vector-like exotics that appear in the 
Orbifold-Landscape.

As discussed next, we extend our feature space by some additional properties of MSSM-like orbifold 
models, in order to obtain a more complex feature space. Moreover, these additional features give a 
notion of MSSM-like models that are more promising from a phenomenological point of view.

\subsection{Heavy top quark from bulk}
\label{sec:HeavyTop}

There is a large hierarchy between the top quark mass and the quark masses of the first and second 
generation. In the MSSM, it is explained by a renormalizable coupling
\begin{equation}\label{eq:heavytop}
\mathcal{W} \supset y_{ij}\,q_i\, \bar{u}_j\, h_u
\end{equation}
in the superpotential $\mathcal{W}$. To first approximation, the matrix of Yukawa couplings 
$y_{ij}$ has to have rank one in order to define the large top quark mass. 

In string theory, the top Yukawa coupling can be related to the ten-dimensional gauge coupling 
constant~\cite{Faraggi:1991be,Faraggi:1995bv,Burdman:2002se,Hosteins:2009xk}. In more detail, from 
a ten-dimensional perspective eq.~\eqref{eq:heavytop} originates from a supersymmetric 
$\E{8}\times\E{8}$ gauge interaction in ten-dimensions. Then, compactifying to four dimensions, the 
coupling eq.~\eqref{eq:heavytop} is present if the left-chiral top quark doublet $q_3$, its 
right-chiral top quark partner $\bar{u}_3$ and the up-type Higgs $h_u$ distribute among the three 
so-called untwisted sectors $U_a$, for $a=1,2,3$, respectively. Here, the untwisted sector $U_a$, 
where $a=1,2,3$, is defined by the (complexified) internal component $A^a\sim A^{2+2a}+\I A^{3+2a}$ 
of the ten-dimensional $\E{8}\times\E{8}$ gauge bosons $A^M$, $M=0,\ldots,9$. This mechanism gives 
an appealing explanation for the large hierarchy in the up-type quark masses and, hence, we append 
our feature space by a feature ``heavy top from bulk''.

\subsection{Non-Abelian flavor symmetries from vanishing Wilson lines}
\label{sec:FlavorSymmetry}

Non-Abelian flavor symmetries of the four-dimensional effective theory can emerge in heterotic 
orbifolds from the localization of certain strings in the extra-dimensions of the orbifold. In more 
detail, string interactions are constrained by so-called string selection rules that describe the 
ability of strings to split, stretch and join while they propagate on the surface of the 
orbifold~\cite{Hamidi:1986vh,Dixon:1986qv}. These constraints can be formulated in terms of Abelian 
discrete symmetries~\cite{Ramos-Sanchez:2018edc}. If certain background fields (i.e.\ Wilson 
lines~\cite{Ibanez:1986tp}) vanish, an additional permutation symmetry of some of the localized 
strings can emerge such that the full flavor symmetry becomes non-Abelian~\cite{Kobayashi:2004ya,Kobayashi:2006wq}. 
In the heterotic orbifold construction, there are two main types of non-Abelian flavor groups for 
MSSM-like orbifold models~\cite{Olguin-Trejo:2018wpw} (being $\Delta(54)$ and $D_{8}$, where $D_8$ 
denotes the dihedral group of order 8, sometimes also denoted by $D_4$). Since non-Abelian flavor 
symmetries are phenomenologically appealing~\cite{Feruglio:2019ktm} and related to a vanishing Wilson 
line in the context of heterotic orbifolds, we extend our feature space by the number of vanishing 
Wilson lines, see also refs.~\cite{Nilles:2014owa,Olguin-Trejo:2018wpw}. In other words, a non-zero 
value of the feature ``\# vanishing Wilson lines'' signals the presence of a non-Abelian flavor 
symmetry.

\subsection{Hidden sector gaugino condensation and supersymmetry breaking}

Supersymmetry breaking through hidden sector gaugino condensation is correlated to the hidden 
sector gauge group and its light matter content~\cite{Nilles:1982ik,Ferrara:1982qs,Derendinger:1985kk,Dine:1985rz}. 
The $\E{8}\times\E{8}$ heterotic string is especially suitable for this mechanism, as it contains, 
beside the observable $\E{8}$ that hosts the MSSM, a hidden $\E{8}$ factor, which generically 
yields supersymmetry breaking at low energies, see refs.~\cite{Lebedev:2006tr,Nilles:2014owa} and 
also~\cite{Dijkstra:2004cc,Dienes:2006ut}. 

For each MSSM-like orbifold model, we compute for each hidden sector non-Abelian gauge group factor 
$G_\mathrm{hidden}$ the chiral part of the spectrum with respect to 
$\SU{3}_C\times\SU{2}_L\times\U{1}_Y\times G_\mathrm{hidden}$ (assuming that the vector-like part 
decouples). Then, the resulting beta-function coefficient is given by 
$b= 3 C_2 - \sum_{\rep{r}} \ell(\rep{r})$ of $G_\mathrm{hidden}$. Here, the summation is performed 
over chiral matter transforming in a representation $\rep{r}$ of $G_\mathrm{hidden}$. Furthermore, 
we have $C_2 = N$ and $\ell(\rep{N}) = \nicefrac{1}{2}$ for 
$\SU{N}$, while we get $C_2 = 2(N-1)$ and $\ell(\rep{2N}) = 1$ for $\SO{2N}$. Then, the gauge 
coupling $g_\mathrm{hidden}(\mu)$ of $G_\mathrm{hidden}$ depends on the energy scale $\mu$, where 
$b$ determines the one-loop RGE, being
\begin{equation}
\frac{\partial g}{\partial\mathrm{ln} \mu} ~=~ -b \frac{g^3}{16\pi^2}\;.
\end{equation}
After solving this differential equation, one can compute the energy scale $\Lambda$ at which the 
coupling $g_\mathrm{hidden}(\mu)$ diverges, i.e.\ when 
$\nicefrac{1}{g_\mathrm{hidden}(\Lambda)} \rightarrow 0$. It is given by 
\begin{equation}\label{eq:Lambda}
\Lambda ~=~ M_\mathrm{GUT}\, \exp\left( -\frac{16 \pi^2}{2\,b\, g^2(M_\mathrm{GUT})}\right)\;,
\end{equation}
with $M_\mathrm{GUT} \approx 3 \cdot 10^{16} \mathrm{GeV}$ and $g^2(M_\mathrm{GUT}) \approx \nicefrac{1}{2}$. 
Furthermore, we assume in eq.~\eqref{eq:Lambda} that the gauge coupling constants of the MSSM 
coincide at the GUT scale $M_\mathrm{GUT}$ approximately with the one of the hidden sector gauge 
group and we neglect string threshold corrections~\cite{Kaplunovsky:1987rp,Dixon:1990pc,Kaplunovsky:1995jw}.
Then, at the energy scale $\Lambda$, the gauginos of $G_\mathrm{hidden}$ form condensates and, 
consequently, supersymmetry is broken spontaneously by a non-vanishing $F$-term of the dilaton. 
Assuming dilaton stabilization by non-perturbative effects and gravity mediation to the observable 
sector, the gravitino mass can be estimated as
\begin{equation}
m_{\nicefrac{3}{2}} ~\approx~ \frac{\Lambda^3}{M^2_\mathrm{Planck}}\;.
\end{equation}
Thus, the feature ``hidden sector beta-function'', specified by the coefficient $b$, gives a rough 
estimate of the scale of supersymmetry breaking.

\begin{figure}[t]
\centering
\includegraphics[scale=0.8]{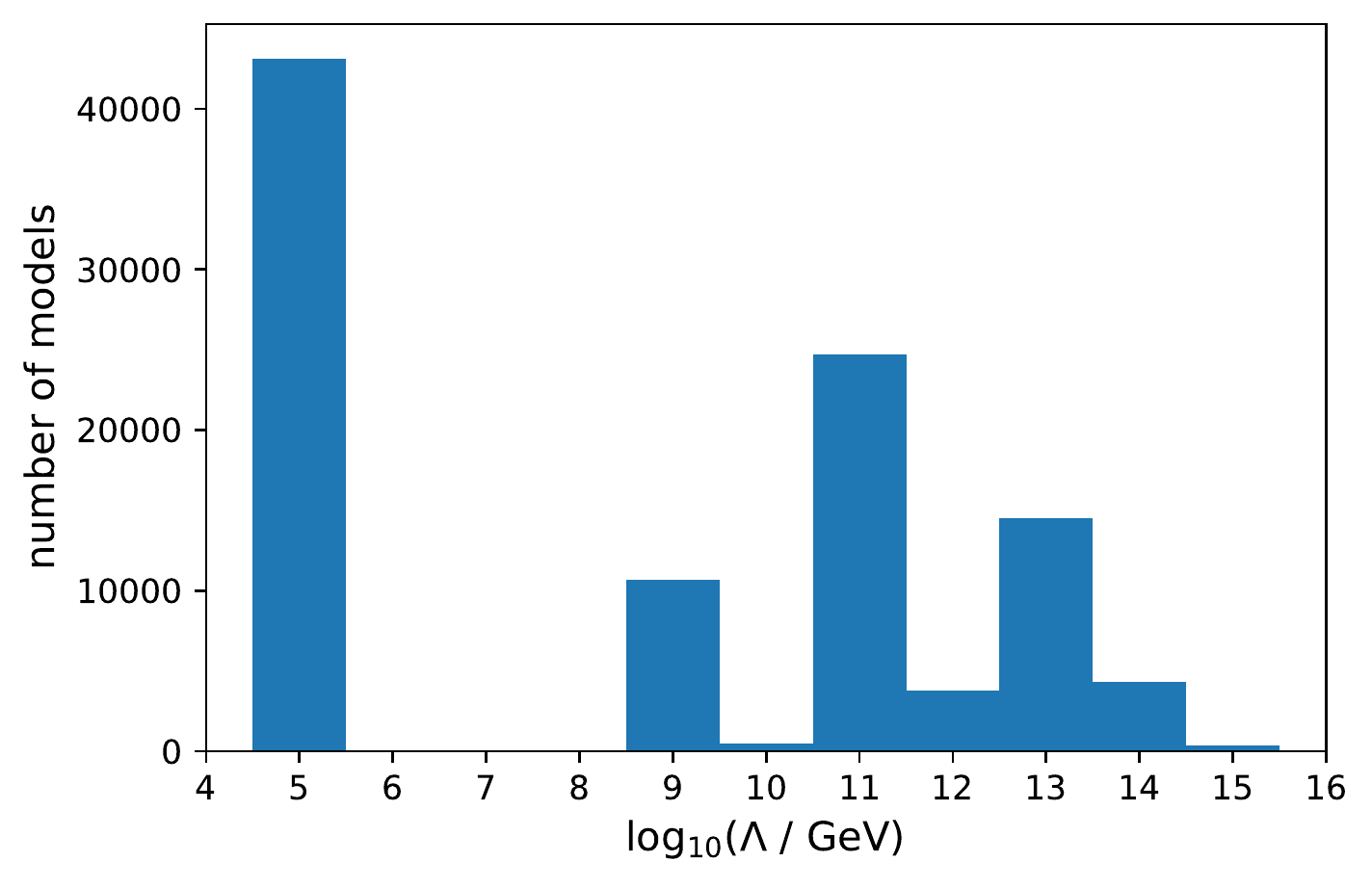}
\vspace{-0.4cm}
\caption{Hidden sector gaugino condensation scale $\Lambda$ for MSSM-like orbifold models based on
all point groups. Note that there are a few models with even smaller $\Lambda$ (and models with 
$b=0$ are excluded). Moreover, the peak at $\Lambda=10^5\, \mathrm{GeV}$ consists mainly of 
MSSM-like orbifold models with point groups $\Z{2}\times\Z{4}$ and $\Z{4}\times\Z{4}$ (each 
contributing $\approx 19,000$ models). For these models, the scale of supersymmetry breaking is far 
too low.\label{fig:Lambda1}}
\end{figure}

\begin{figure}[t]
\centering
\includegraphics[scale=0.8]{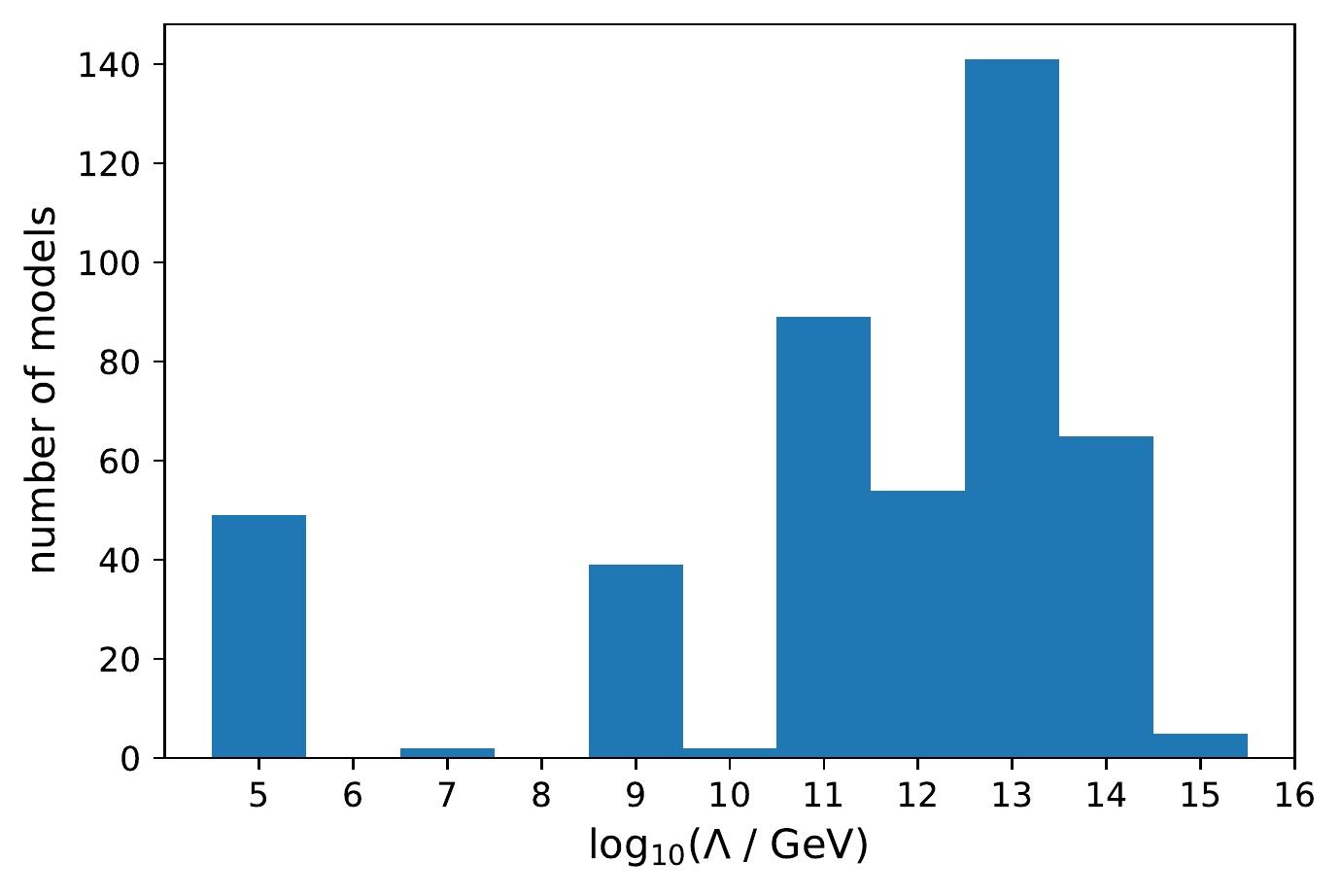}
\vspace{-0.4cm}
\caption{Hidden sector gaugino condensation scale $\Lambda$ for MSSM-like orbifold models based on 
the $\Z6$-$\mathrm{II}$ $(1,1)$ orbifold geometry (models with $b=0$ are excluded), see~\cite{Lebedev:2006tr}.\label{fig:Lambda2}}
\end{figure}

In our MSSM-like orbifold models, the distribution of $\Lambda$ is given in 
figure~\ref{fig:Lambda1} for all point groups and in figure~\ref{fig:Lambda2} for $\Z{6}$-II only. 
We are interested in the case of supersymmetry breaking around (at least) the TeV scale. A very 
rough estimate gives the constraint $\Lambda\gtrsim 10^{13\pm 1}\, \mathrm{GeV}$~\cite{Lebedev:2006tr}, 
which translates to $b \gtrsim 15$.

As a remark, the beta-function coefficient $b$ is closely related to the number of unbroken roots 
$N_{\textrm{ur}}$, i.e.\ the number of roots from the hidden $\E{8}$ factor that survives the 
orbifold projection conditions. Interestingly, the contrast patterns developed in 
ref.~\cite{Parr:2019bta} showed that large values of $N_{\textrm{ur}}$ correlate with a higher 
production rate of MSSM-like orbifold models. Moreover, there are many MSSM-like orbifold models 
with large values of $N_{\textrm{ur}}$ that are (practically) inaccessible by traditional search 
algorithms and were uncovered recently in ref.~\cite{Parr:2019bta}.

\subsection[Anomalous U(1)]{\boldmath Anomalous $\U{1}$\unboldmath}
\label{sec:AnomalousU1}

99\% of the MSSM-like orbifold models have an additional $\U{1}_{\mathrm{anom}}$ gauge factor 
beside hypercharge that appears to be anomalous, where the anomaly is canceled by a universal 
Green--Schwarz mechanism~\cite{Green:1984sg}. This induces a Fayet--Iliopoulos $D$-term (FI-term), 
which sets the scale for a Froggatt--Nielsen-like mechanism~\cite{Froggatt:1978nt}, where some SM 
singlets $s^0$ develop non-vanishing VEVs $\langle s^0\rangle$ in order to satisfy $D_{\mathrm{anom}}=0$. 
Consequently, these VEVs spontaneously break the additional $\U{1}$ factors, generate masses of the 
lighter quarks and leptons, and decouple (at least some of) the vector-like exotics, see 
section~\ref{sec:vectorlikeexotics}. Hence, the existence of an ``anomalous $\U{1}$'' is a good 
feature that characterizes promising orbifold models.

\subsection{Comments on our feature space}

This concludes our feature space. Let us remark that our dataset of $126,783$ inequivalent MSSM-like 
orbifold models corresponds to a set of $106,027$ inequivalent feature vectors, which we 
call $\mathbbm{D}$. The main reason for the decrease from $126,783$ to $106,027$ is associated to 
the hidden sector: In order to distinguish between inequivalent MSSM-like orbifold models, the 
observable and the hidden sector are taken into account~\cite{Nilles:2011aj}. On the other hand, 
the feature vector is supposed to characterize only the properties of a given model with respect to 
the MSSM. Thus, models with different hidden sectors can yield the same feature vector. Taking this 
into account, our features give a good measure to distinguish between inequivalent phenomenological 
properties of MSSM-like orbifold models.

\section{Boosted decision tree}
\label{sec:Boosteddecisiontree}

A boosted decision tree is build up by an ensemble of single decision trees. Each decision tree is 
a so-called weak learner: on its own, it typically yields a poor performance. However, a weighted 
majority vote of many weak learners tend to perform much better. 

In more detail, the idea of boosting is that one combines the predictions of many weak learners 
(e.g.\ decision trees) to get a more powerful estimator. This is achieved by a successive training 
of many weak learners, where the misclassified training data of the previous weak learner is weighted 
with a higher value for the next weak learner in order to enforce him to classify these data points 
correctly (and the weights of the data that has been classified correctly are decreased accordingly). 
In our case, this procedure is repeated 1,500 times. Finally, we combine the 
individual decision trees to a much more powerful estimator: the boosted decision tree. For further 
details on boosting, see for example ref.~\cite{hastie01statisticallearning}.

\subsection{How to measure the performance of ML models for imbalanced datasets}
\label{sec:ml_model_selection}

The performance of a predictive ML model can be measured by the accuracy that is defined by the 
number of correct predictions divided by the total number of all predictions. However, for an 
imbalanced dataset, as the one we have in our case, this normal accuracy measurement can be misled 
in the following way: Assume a classification task with two classes A and B, where the class A 
builds the majority of the dataset with 99\% of all instances. Any prediction method can now 
achieve an accuracy of 99\% simply by predicting class A always, but never class B. 

In order to avoid such a behavior, there exist several metrics for imbalanced classification tasks. 
The one we are using is based on three types of predictions involving the point group $P_i$, i.e.\ 
for each point group $P_i$ we define
\begin{eqnarray}\label{eq:metrics}
\#\ \mathrm{of}\ \mathrm{true}\ \mathrm{positives}  & : & \mathrm{TP}_i ~:=~ \big|\{X \in \mathbbm{D} ~|~ Y_\mathrm{correct}(X)= P_{i} \quad\mathrm{and}\quad Y_\mathrm{predicted}(X)=P_{i}\}\big|\;,\nonumber\\
\#\ \mathrm{of}\ \mathrm{false}\ \mathrm{positives} & : & \mathrm{FP}_i ~:=~ \big|\{X \in \mathbbm{D} ~|~ Y_\mathrm{correct}(X)= P_{j} \quad\mathrm{but}\quad Y_\mathrm{predicted}(X)=P_{i}\}\big|\;,\nonumber\\
\#\ \mathrm{of}\ \mathrm{false}\ \mathrm{negatives} & : & \mathrm{FN}_i ~:=~ \big|\{X \in \mathbbm{D} ~|~ Y_\mathrm{correct}(X)= P_{i} \quad\mathrm{but}\quad Y_\mathrm{predicted}(X)=P_{j}\}\big|\;.\nonumber
\end{eqnarray}
Here, the indices $i \neq j$ label the 14 different orbifold point groups $P_i$, $X$ is a feature 
vector from our dataset $\mathbbm{D}$, and $Y_\mathrm{correct}(X)$ 
$\big(Y_\mathrm{predicted}(X)\big)$ label the correct (predicted) point group of $X$, respectively. 
Counting for each point group $P_i$ the numbers $\mathrm{TP}_i$, $\mathrm{FP}_i$ and $\mathrm{FN}_i$ 
allows us to define three different metrics,
\begin{subequations}
\begin{eqnarray}
\mathrm{precision}               &:&\qquad \mathrm{p}_i  ~:=~   \dfrac{\mathrm{TP}_i}{\mathrm{TP}_i + \mathrm{FP}_i}\;,  \\
\mathrm{recall}                  &:&\qquad \mathrm{r}_i  ~:=~   \dfrac{\mathrm{TP}_i}{\mathrm{TP}_i + \mathrm{FN}_i}\;,  \\
\mathrm{f1}\text{-}\mathrm{score}&:&\qquad \mathrm{f1}_i ~:=~ 2 \dfrac{\mathrm{p}_i\; \mathrm{r}_i}{\mathrm{p}_i + \mathrm{r}_i}\;. 
\end{eqnarray}
\end{subequations}
Finally, one can define the ``f1-macro'' as the average of the f1-scores for all point groups. 
Then, to deal with our highly imbalanced data, we use the f1-macro to measure the performance of our 
ML model.

Let us briefly illustrate the benefit of using the f1-macro on the example from the beginning of 
this section. In this case, we have $\mathrm{f1}_A \approx 1$ and $\mathrm{f1}_B = 0$ (for the 
extreme case of predicting always $A$ and never $B$, the precision value of $B$ is actually 
undefined. However, one can easily rewrite the f1-score directly in terms of 
$\mathrm{TP}_B=\mathrm{FP}_B=0$ and $\mathrm{FN}_B \neq 0$ in order to see that 
$\mathrm{f1}_B = 0$). Consequently, the f1-macro is given by 
$\nicefrac{1}{2}(\mathrm{f1}_A+\mathrm{f1}_B) \approx 0.5$, which rates the naive classification 
as insufficient.

\subsection{Training of the ML model}

We start by splitting our dataset $\mathbbm{D}$ and the corresponding target values 
$Y_\mathrm{correct}$, i.e.\ the orbifold point groups $P_{i}$, of our $106,027$ feature vectors 
into 80\% training and 20\% test data. The test set $\mathbbm{T}$ is held back for the evaluation 
of the trained ML model. On the other hand, the training set is used to perform a grid search for 
the optimal hyperparameters, i.e.\ each set of hyperparameters in the grid is used to train an ML 
model using 5-fold cross-validation (\texttt{CV}). Then, the best ML model is chosen based on the 
f1-macro. In more detail, the module \texttt{GridSearchCV} from the scikit-learn 
library~\cite{scikit-learn} is utilized for hyperparameter search and the LightGBM 
implementation~\cite{Ke2017LightGBMAH} for boosted decision trees.

After an extensive hyperparameter search, it turns out that the following hyperparameters give the 
best f1-macro performance:
\begin{itemize}
\item \texttt{class\_weight=`balanced':}
The argument ``balanced'' weights the 14 classes of orbifold point groups inversely proportional to 
their frequency of occurrence in the input dataset at the beginning of the training. In detail, 
the dominant classes (like $\Z{2}\times\Z{4}$ and $\Z{4}\times\Z{4}$) get weights smaller than 1 
and the small classes (like $\Z{4}$) get weights larger than 1.
\item \texttt{learning\_rate=0.2} 
\item \texttt{min\_child\_samples=8:} 
The minimal number of samples per child node, where the default value is 20. This is beneficial 
since there are nodes with a small number of samples.
\item \texttt{min\_child\_weight=0.01:}
A regularization measure to stop splitting a node if its purity is high.
\item \texttt{n\_estimators=1500:} 
The number of individual decision trees.
\item \texttt{num\_leaves=50:} 
The maximum number of leaf nodes for each decision tree.
\end{itemize}
To evaluate the predictive power of the final ML model with optimized hyperparameters, we use the 
test set $\mathbbm{T}$. The results of this evaluation will be discussed in 
section~\ref{sec:Performance}.

\begin{figure}[h!]
\centering
\includegraphics[width=\textwidth]{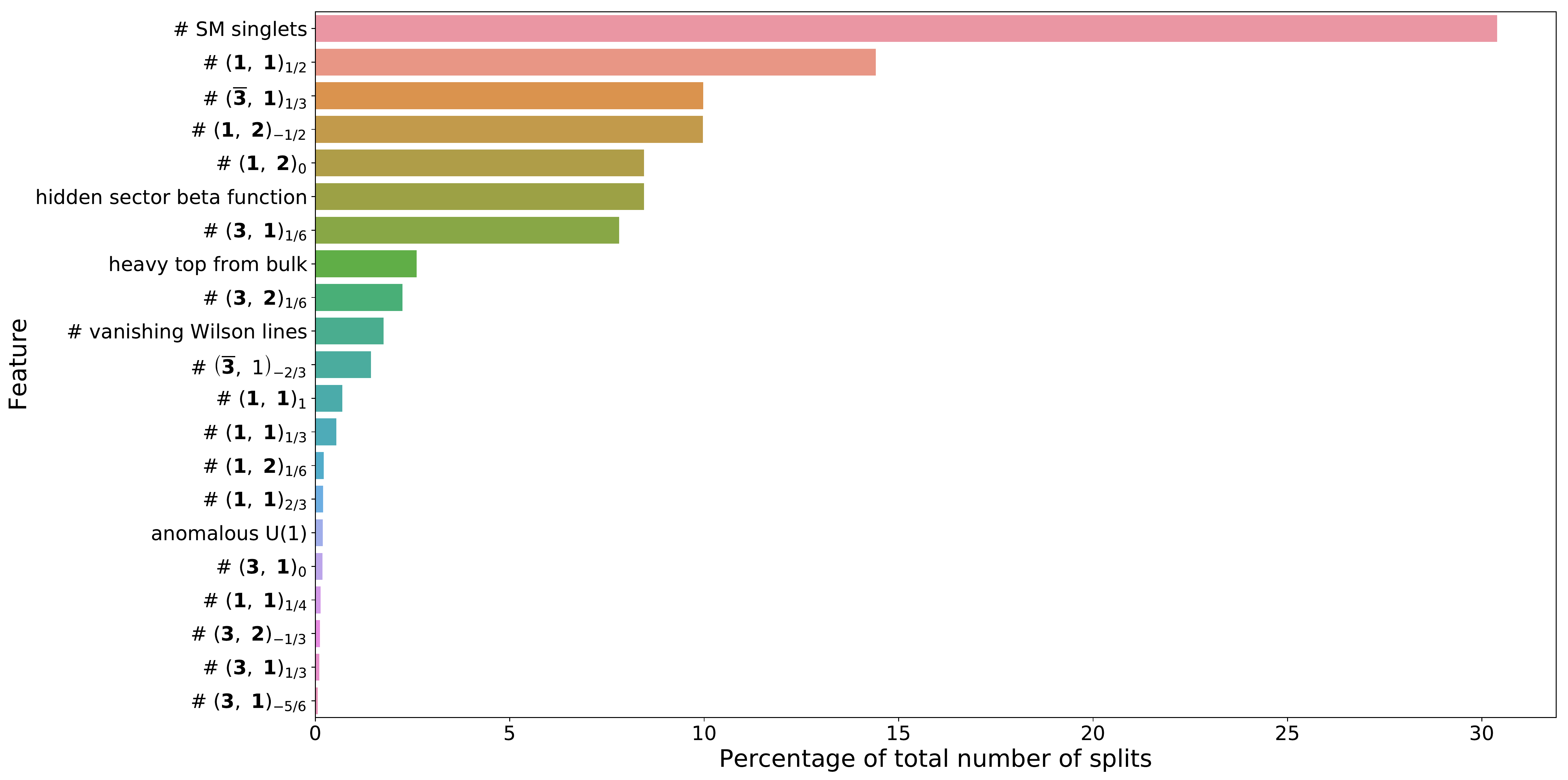}
\caption{\label{fig:FeaturesInDecisionTree}
The importance of the features (defined in section~\ref{sec:OrbifoldLandscape}) for the 
classification task performed by our boosted decision tree: For each decision tree of the ensemble 
of trees, we count the number how often each feature was used in the nodes of the tree. Then, for 
each feature we take the average over all trees. Finally, we give the numbers as percentage and 
rank the features accordingly. As a result, the feature ``\# SM singlets'' turns out to be used in 
30\% of the nodes and is the most important feature for our boosted decision tree.}
\end{figure}

In addition to a boosted decision tree (i.e.\ LightGBM), we also tried various alternative 
classification algorithms. To be specific: k-nearest-neighbors, linear and quadratic discriminant 
analysis, logistic regression, random forest, support vector machines, and fully connected neural 
networks with softmax classification.\footnote{The implementations from scikit-learn~\cite{scikit-learn} 
are used for the non-neural network algorithms, whereas the Keras API~\cite{chollet2015keras} is 
used for the neural networks.} Non of those alternative ML models individually performed on 
the level of LightGBM. Only XGBoost~\cite{DBLP:journals/corr/ChenG16}, a different implementation 
of boosted decision trees, yields comparable results. In addition, we build an ensemble of these 
different estimators, where the prediction of the ensemble is a weighted linear combination of the 
predictions of each individual estimator. However only a combination of LightGBM with different 
neural network architectures shows a slight improvement, where the f1-macro increases from 0.856 for 
LightGBM as a single classifier to 0.867 for the ensemble. Since this improvement is small and our 
main results do not change, we decide to keep things simple and use LightGBM as a single estimator 
only. In addition, the usage of LightGBM as a single estimator allows us to read out and interpret 
the inner structure of the boosted decision tree by visualizing the feature importance, see 
figure~\ref{fig:FeaturesInDecisionTree}.

\begin{table}[t]
\centering
\begin{tabular}{c|ccc|c}
point group         & precision       &  recall         & f1-score        & \multirow{2}{*}{support} \\
$P_i$               & $\mathrm{p}_i$  &  $\mathrm{r}_i$ & $\mathrm{f1}_i$ &  \\
\hline
\hline
   $\Z{4}$          &    0.85   &   0.85   &   0.85   &     33     \\
   $\Z{6}$-I        &    0.88   &   0.88   &   0.88   &      8    \\
   $\Z{6}$-II       &    0.82   &   0.75   &   0.79   &    305    \\
   $\Z{8}$-I        &    0.75   &   0.78   &   0.76   &    125    \\
   $\Z{8}$-II       &    0.86   &   0.84   &   0.85   &    444    \\
  $\Z{12}$-I        &    0.95   &   0.90   &   0.92   &    232    \\
  $\Z{12}$-II       &    0.69   &   0.68   &   0.69   &     82     \\
$\Z{2}\times\Z{2}$  &    0.88   &   0.83   &   0.85   &    255     \\
$\Z{2}\times\Z{4}$  &    0.97   &   0.98   &   0.98   &  8,155     \\
$\Z{2}\times\Z{6}$-I&    0.84   &   0.59   &   0.69   &    211    \\
$\Z{3}\times\Z{3}$  &    0.99   &   1.00   &   0.99   &    483     \\
$\Z{3}\times\Z{6}$  &    0.87   &   0.88   &   0.87   &  1,019    \\
$\Z{4}\times\Z{4}$  &    0.98   &   0.98   &   0.98   &  9,113     \\
$\Z{6}\times\Z{6}$  &    0.93   &   0.96   &   0.94   &    741    \\
\hline
\hline
average        &    0.88   &   0.85   &   0.86   &  21,206    \\
\end{tabular}
\caption{Classification report of our boosted decision tree, evaluated on the test set 
$\mathbbm{T}\subset\mathbbm{D}$ of $21,206$ feature vectors of MSSM-like orbifold models, see 
eq.~\eqref{eq:metrics} for definitions. The f1-macro (i.e.\ the average of f1-scores) is $0.86$. 
The support for each point group $P_i$ is defined as 
$\big|\{X \in \mathbbm{T} ~|~ Y_\mathrm{correct}(X)= P_{i}\}\big|$. The best performance is achieved 
for MSSM-like orbifold models with point groups $\Z{3}\times\Z{3}$, $\Z{4}\times\Z{4}$ and 
$\Z{2}\times\Z{4}$.\label{tab:classification_report}}
\end{table}

\begin{table}[h!]
\centering
\vspace{2cm}
\resizebox{\textwidth}{!}{
\begin{tabular}{rr||cccccccccccccc}
\parbox[t]{2mm}{\multirow{18}{*}{\rotatebox[origin=c]{90}{\Large{correct orbifold point group}}}} & \multicolumn{14}{c}{\Large{predicted orbifold point group}} \\
                 &           & $\Z{4}$ & $\Z{6}$-I & $\Z{6}$-II & $\Z{8}$-I & $\Z{8}$-II   & $\Z{12}$-I & $\Z{12}$-II & $\Z{2}\times\Z{2}$ & $\Z{2}\times\Z{4}$ & $\Z{2}\times\Z{6}$-I & $\Z{3}\times\Z{3}$ & $\Z{3}\times\Z{6}$ &  $\Z{4}\times\Z{4}$ & $\Z{6}\times\Z{6}$   \\
\hline \hline
    & $\Z{4}$               &   28    &     0     &     0      &      1    &     0        &     0      &      0      &     0              &        4           &    0                 &     0              &     0              &      0              &     0                \\
    & $\Z{6}$-I             &    0    &     7     &     0      &      0    &     0        &     0      &      0      &     0              &        0           &    0                 &     0              &     0              &      1              &     0                \\
    & $\Z{6}$-II            &    0    &     0     &   229      &      0    &    14        &     0      &      4      &     0              &       55           &    2                 &     1              &     0              &      0              &     0                \\
    & $\Z{8}$-I             &    1    &     0     &     0      &     97    &     1        &     2      &      0      &     0              &        0           &    0                 &     0              &     1              &     23              &     0                \\
    & $\Z{8}$-II            &    0    &     0     &    15      &      0    &   375        &     0      &     10      &     2              &       41           &    0                 &     0              &     0              &      1              &     0                \\
    & $\Z{12}$-I            &    0    &     0     &     0      &      4    &     0        &   209      &      0      &     0              &        0           &    0                 &     2              &     0              &     17              &     0                \\
    & $\Z{12}$-II           &    0    &     0     &     4      &      0    &    10        &     0      &     56      &     0              &       12           &    0                 &     0              &     0              &      0              &     0                \\
    & $\Z{2}\times\Z{2}$    &    0    &     0     &     0      &      0    &     2        &     0      &      0      &   212              &       41           &    0                 &     0              &     0              &      0              &     0                \\
    & $\Z{2}\times\Z{4}$    &    3    &     0     &    24      &      4    &    31        &     1      &      7      &    27              &    8,004           &   21                 &     0              &     5              &     26              &     2                \\
    & $\Z{2}\times\Z{6}$-I  &    0    &     0     &     2      &      1    &     3        &     1      &      3      &     0              &       69           &  124                 &     0              &     2              &      4              &     2                \\
    & $\Z{3}\times\Z{3}$    &    0    &     0     &     0      &      0    &     0        &     2      &      0      &     0              &        0           &    0                 &   481              &     0              &      0              &     0                \\
    & $\Z{3}\times\Z{6}$    &    0    &     0     &     1      &      0    &     2        &     0      &      0      &     0              &        0           &    0                 &     0              &   899              &     93              &    24                \\
    & $\Z{4}\times\Z{4}$    &    1    &     1     &     3      &     22    &     0        &     5      &      1      &     0              &       21           &    1                 &     0              &   116              &   8,914              &    28                \\
    & $\Z{6}\times\Z{6}$    &    0    &     0     &     0      &      0    &     0        &     0      &      0      &     0              &        0           &    0                 &     0              &    16              &     12              &   713                \\
\end{tabular}
}
\caption{Confusion matrix of our boosted decision tree, evaluated on the test set 
$\mathbbm{T}\subset\mathbbm{D}$ of $21,206$ feature vectors of MSSM-like orbifold models.}
\label{tab:confusion_matrix}
\end{table}

\subsection{Performance of the ML model}
\label{sec:Performance}

Next, we analyze the performance of our trained ML model on the test set $\mathbbm{T}$. This is 
quantified in a classification report, see table~\ref{tab:classification_report}. We see that for 
all orbifold point groups (even for the minority classes like $\Z{4}$ and $\Z{6}$-I) the f1-score 
is very high. In addition, we also state the confusion matrix in table~\ref{tab:confusion_matrix}. 
These results indicate that our boosted decision tree is well balanced. It is 
intriguing how well our ML model can predict the orbifold point group using only the spectrum of 
vector-like exotics and some additional phenomenologically appealing features. Note also that the 
MSSM-like orbifold models from the training set are only unique among a certain orbifold point 
group. In some cases, MSSM-like orbifold models from different point groups yield similar 
(or even identical) feature vectors. Then, the classifier has to decide to which orbifold point 
group this part of the feature space consists more likely. This introduces some intrinsic 
uncertainty to our ML model and some misclassifications are unavoidable

\newpage

\begin{figure}[t]
\centering
\includegraphics[scale=0.4]{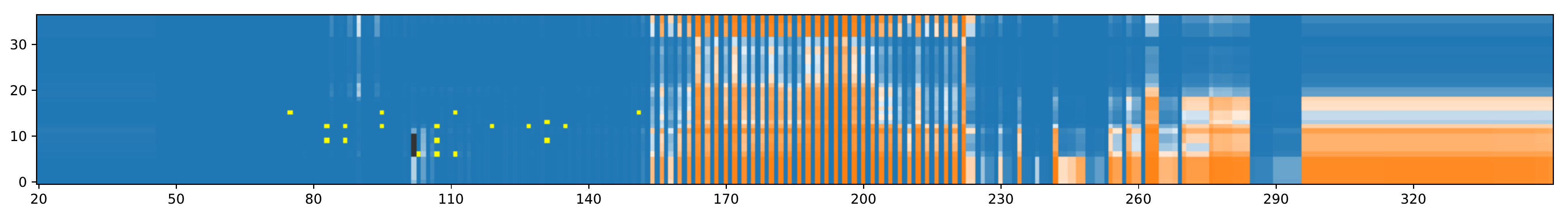}
\caption{Stringy origin of the MSSM with ``heavy top from bulk'' but without a non-Abelian flavor 
symmetry. The colors are associated to point groups as follows: blue for $\Z{2}\times\Z{4}$, orange 
for $\Z{4}\times\Z{4}$ and black for $\Z{8}$-$\mathrm{I}$. In yellow, we highlight 20 MSSM-like orbifold models 
that have no vector-like exotics beside some Higgs-pairs, see section~\ref{sec:AlmostPerfectMSSM}. 
See also figure~\ref{fig:pred_MSSM} for additional cases and further details.
\label{fig:pred_almost_perfect_MSSM}
}
\end{figure}

\section{Predicting the stringy origin of the MSSM}
\label{sec:results}

After we have shown that the predictions of our classifier are reasonable for the given feature 
vectors of MSSM-like orbifold models, the next step is to extrapolate from these results towards the 
MSSM: we give a feature vector without any vector-like exotics to the ML algorithm in order to 
identify its most likely origin from orbifold compactifications. Extrapolations with ML models are 
in general rather difficult: we try to make a prediction for a feature vector that is different to 
the data from $\mathbbm{D}$. However, since we use additional features beside the numbers of 
vector-like exotics (like the number of SM singlets and the hidden sector beta-function 
coefficient, see section~\ref{sec:OrbifoldLandscape}), the prediction for the MSSM also includes 
non-trivial features that are embedded in the orbifold dataset. In other words, the feature vector 
of the MSSM gets closer to the feature vectors of our MSSM-like orbifold models by using these 
additional features. Moreover, our experience with regularized decision trees (see 
ref.~\cite{Parr:2019bta}) indicates that they are suitable for extrapolations: In general, each 
decision tree divides up the feature space into smaller subspaces and assigns a prediction (i.e.\ 
an orbifold point group) to each subspace. However, a decision tree necessarily leaves subspaces of 
infinite volume at the boundary of the training set. Since the feature vector of the MSSM is 
outside the region of feature vectors of MSSM-like orbifold models, the MSSM will lie in one of 
these infinite subspaces, but still gets a meaningful prediction assigned. Then, an ensemble of 
decision trees gives an additional regularization that improves the generalization to a wider range 
in feature space. In this way, boosted decision trees can be used to get meaningful predictions, even 
for extrapolations.

\begin{figure}[t!]
\centering
  \subfloat[Stringy origin of the MSSM: with heavy top from bulk and a non-Abelian flavor 
            symmetry.\label{fig:pred_hT1_WL1}]
            {\includegraphics[scale=0.4]{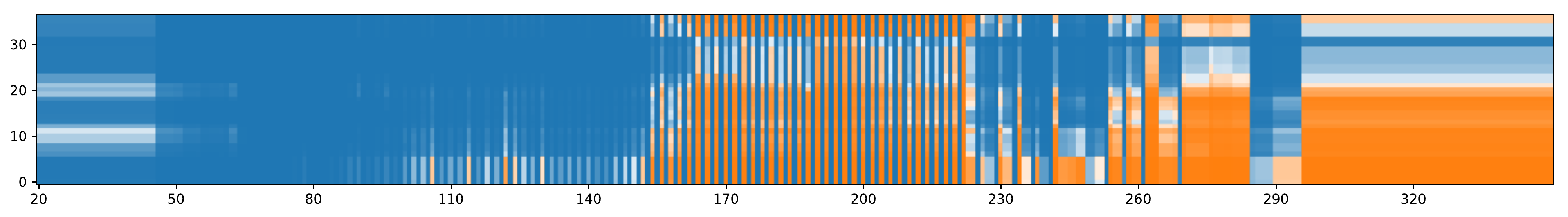}}\\

  \subfloat[Stringy origin of the MSSM: with heavy top from bulk but without a non-Abelian flavor 
            symmetry. The small black region for an MSSM with $\approx 100$ SM singlets corresponds 
            to $\Z{8}$-$\mathrm{I}$. \label{fig:pred_hT1_WL0}]
            {\includegraphics[scale=0.4]{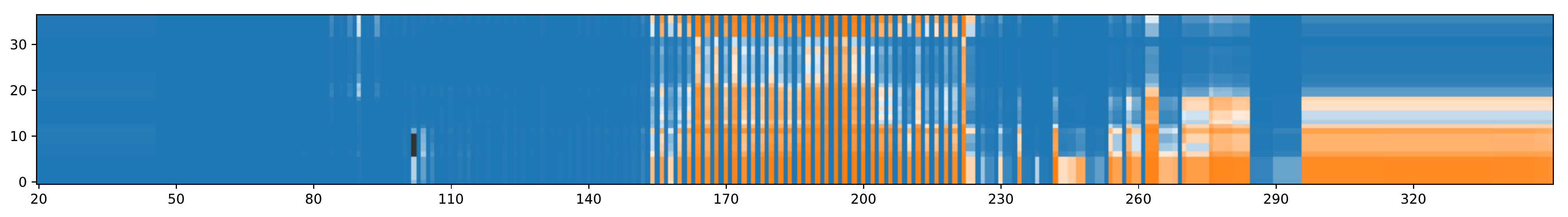}}\\
            
  \subfloat[Stringy origin of the MSSM: without heavy top from bulk but with a non-Abelian flavor 
            symmetry. The small black regions correspond to either $\Z{6}$-$\mathrm{II}$ or 
            $\Z{8}$-$\mathrm{I}$.\label{fig:pred_hT0_WL1}]
            {\includegraphics[scale=0.4]{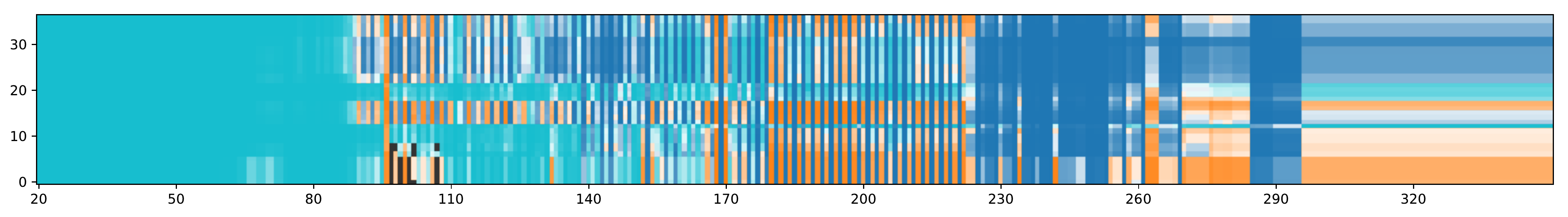}}\\
                        
  \subfloat[Stringy origin of the MSSM: without heavy top from bulk and without a non-Abelian 
            flavor symmetry. The black regions correspond to either $\Z{12}$-$\mathrm{I}$, 
            $\Z{3}\times\Z{6}$, $\Z{6}$-$\mathrm{I}$, $\Z{6} \times \Z{6}$ or 
            $\Z{8}$-$\mathrm{I}$.\label{fig:pred_hT0_WL0}]
            {\includegraphics[scale=0.4]{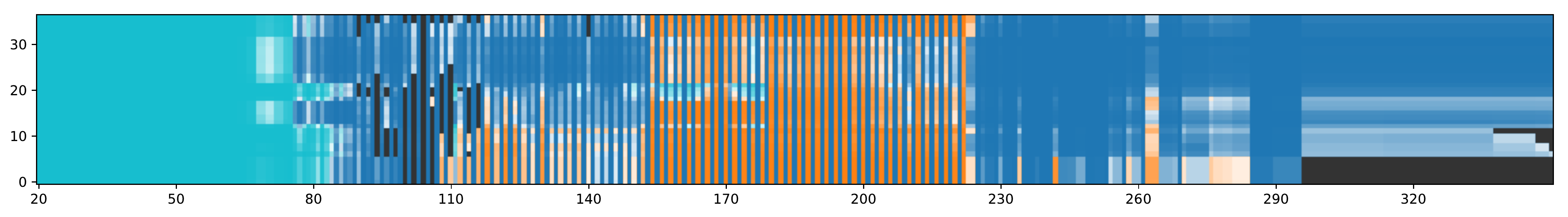}}\\
            
  \subfloat{\hspace{0.05cm} \includegraphics[scale=0.4]{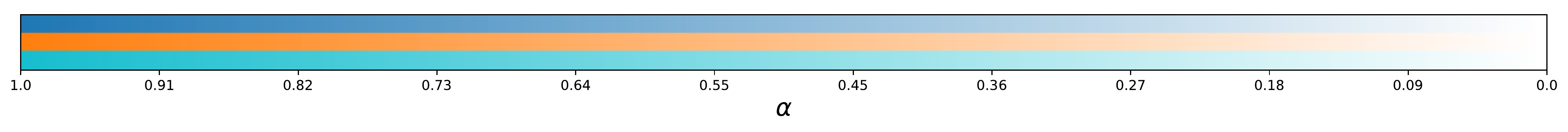}}\\
\caption{
\label{fig:pred_MSSM}
Predicted origin of the MSSM for four different cases (a) - (d) depending on the features ``heavy 
top from bulk'', see section~\ref{sec:HeavyTop} and ``non-Abelian flavor symmetry'', see 
section~\ref{sec:FlavorSymmetry}: in each case, the x-axis corresponds to the number of SM singlets 
(i.e.\ right-handed neutrinos), while the y-axis gives the beta-function coefficient $b$ that sets 
the scale of supersymmetry breaking via hidden sector gaugino condensation. \\ The colors are 
associated to point groups as follows:\\ \hspace*{0.3cm}(i) blue: $\Z{2}\times\Z{4}$,\\\hspace*{0.3cm}(ii) 
orange: $\Z{4}\times\Z{4}$,\\\hspace*{0.3cm}(iii) turquoise: $\Z{2}\times\Z{2}$.\\
Black indicates 
a different point group than the three dominant ones. The transparency 
$\alpha:=p(1^\mathrm{st}) - \tfrac{3}{4} p(2^\mathrm{nd})$ of each pixel indicates the difference 
between the highest and the second highest probabilities of the predictions, $p(1^\mathrm{st})$ and 
$p(2^\mathrm{nd})$, respectively. Note that $\sum_{i} p(P_{i}) = 1$. Hence, the color gets very 
transparent if $p(1^\mathrm{st}) \approx p(2^\mathrm{nd})$.}
\end{figure}

In detail, in order to predict the orbifold origin of the MSSM, we generate a feature vector for 
each value of the hidden sector beta-function coefficient $b \in\{0,\ldots,36\}$ (i.e.\ between the 
minimal and maximal value of $b$ in our dataset $\mathbbm{D}$) and for each number of Standard 
Model singlets: \# SM singlets $\in\{20, \dots, 350\}$. In this way, we obtain $37\cdot 331$ feature 
vectors that we give to our trained ensemble of decision trees in order to obtain a prediction for 
each of them. We can illustrate the results in a two--dimensional plot, where different colors correspond to 
different orbifold point groups and the transparency of the color indicates the degree of accuracy 
of the corresponding prediction, see figure~\ref{fig:pred_almost_perfect_MSSM} and the caption of 
figure~\ref{fig:pred_MSSM} for further details.

\clearpage

Furthermore, we turn on and off the properties of non-Abelian flavor symmetries (corresponding to 
the ``\# vanishing Wilson lines'') as well as ``heavy top from bulk''.  In this way, we obtain 
three additional cases and we display for each case the predictions of the $37\cdot331$ feature 
vectors in figure~\ref{fig:pred_MSSM}.
It turns out that in general the predictions are dominated by two classes: $\Z2\times\Z4$ and 
$\Z4\times\Z4$. Interestingly, these classes build up the majority of MSSM-like orbifold models 
with $55,429$ and $48,812$ MSSM-like models, respectively, and they achieve the highest f1-scores, 
see table~\ref{tab:classification_report}. In addition, for MSSMs without a heavy top from the 
bulk, the $\Z2\times\Z2$ orbifold point group (with $1,711$ MSSM-like models) occupies 
a large fraction of the prediction area, especially in cases with less than $\approx 80$ SM 
singlets, see figures~\ref{fig:pred_hT0_WL1} and~\ref{fig:pred_hT0_WL0}. Finally, for MSSMs without 
a heavy top from the bulk and without non-Abelian flavor symmetries, several orbifold point groups 
become relevant in distinct areas of figure~\ref{fig:pred_hT0_WL0}: for MSSMs with $90$ to $115$ SM 
singlets the $\Z{8}$-I point group appears, while for MSSMs with large $b$ and $\approx 90$ or 
$140$ SM singlets the point groups $\Z{12}$-I or $\Z{6}$-I get predicted, respectively. Moreover, 
MSSMs with many ($\approx 300$) SM singlets are predicted to originate from orbifolds with 
$\Z{3}\times\Z{6}$ or $\Z{6} \times \Z{6}$ point groups for $b \lesssim 5$ or 
$5 \lesssim b \lesssim 10$, respectively.

\subsection{An almost perfect MSSM-like orbifold model}
\label{sec:AlmostPerfectMSSM}

In this section, we present an explicit ``almost perfect'' MSSM-like orbifold model, based on the 
point group $\Z{2}\times\Z{4}$ (i.e.\ the orbifold geometry is labeled $\Z{2}\times\Z{4}$ $(2, 4)$ 
in the nomenclature of ref.~\cite{Fischer:2012qj}; also see ref.~\cite{Pena:2012ki} for string 
model building based on the $\Z{2}\times\Z{4}$ orbifold). In this case, the six-torus is 
non-factorizable and both rotational generators of the orbifold are 
roto-translations, defined as
\begin{equation}
\left(\theta, \frac{1}{2}\left(e_1+e_2+e_6\right)\right) \quad\mathrm{and}\quad \left(\omega, \frac{1}{2}\left(e_1+e_2+e_6\right)\right)\;,
\end{equation}
where $\theta^2 = \omega^4 = \Id$. The $\Z{2}\times\Z{4}$ shift vectors associated to $\theta$ and 
$\omega$ are chosen as
\begin{subequations}
\begin{eqnarray}
  V_1 & = & \left(    0,     0,     0,     0, \tfrac{1}{2}, \tfrac{1}{2}, \tfrac{3}{2}, \tfrac{3}{2}\right)  \left(    0, \tfrac{1}{2}, \tfrac{1}{2}, \tfrac{1}{2}, \tfrac{1}{2}, \tfrac{1}{2}, \tfrac{1}{2},     2\right)\;,\\
  V_2 & = & \left(-\tfrac{1}{4}, \tfrac{1}{4}, \tfrac{1}{4},     1,    -1,     0,    -1, \tfrac{1}{4}\right)  \left(-\tfrac{3}{4},     0,     0,     0,     0,     0, \tfrac{1}{2}, -\tfrac{3}{4}\right)\;,
\end{eqnarray}
\end{subequations}
respectively. Furthermore, the non-trivial Wilson lines $W_i$ associated to the six independent 
directions $e_i$ of the compact orbifold space read
\begin{subequations}
\begin{eqnarray}
  W_{3} & = & \left(-\tfrac{1}{4}, \tfrac{1}{4}, \tfrac{5}{4}, -\tfrac{5}{4}, \tfrac{5}{4}, \tfrac{5}{4}, -\tfrac{5}{4}, -\tfrac{5}{4}\right)  \left(-\tfrac{3}{4}, -\tfrac{5}{4}, -\tfrac{1}{4}, -\tfrac{1}{4}, \tfrac{1}{4}, \tfrac{5}{4}, \tfrac{9}{4}, \tfrac{3}{4}\right)\;,\\
  W_{5} & = & \left(    0,     0,     0,     0,     0,     0,     0,     0\right)  \left(\tfrac{3}{2}, -\tfrac{1}{2},    -1,     2,     0, -\tfrac{3}{2},     1, \tfrac{1}{2}\right)\;,
\end{eqnarray}
\end{subequations}
while $W_{3} = W_{4} = W_{6}$ and $W_{1} = W_{2} = (0^{16})$ are fixed due to geometric constraints. 
These shifts and Wilson lines act on the $\E{8}\times\E{8}$ gauge degrees of freedom. Using the 
orbifolder~\cite{Nilles:2011aj}, the resulting four-dimensional gauge group reads
\begin{equation}
\SU{3}_\mathrm{flavor} \times \SU{3}_\mathrm{C} \times \SU{2}_\mathrm{L} \times \SU{5}_\mathrm{hidden} \times \U{1}_\mathrm{Y} \times \U{1}^6\;,
\end{equation}
where $\SU{3}_\mathrm{flavor}$ is a gauged $\SU{3}$ flavor symmetry and one of the $\U{1}$'s 
appears anomalous. Gauge symmetry breaking is non-local in this model, i.e.\ it is associated to a 
non-trivial fundamental group of the orbifold with a Wilson line that breaks an intermediate 
$\SU{5}$ GUT to the SM. Consequently, there exist stable winding strings (with masses related to 
the compactification scale) which can contribute to the dark matter relic density~\cite{Mutter:2019byx}. 
Finally, the massless matter spectrum is given in table~\ref{tab:AlmostPerfectMSSM}. From the point 
of view of the MSSM, this model contains exactly three generations of quarks and leptons, three 
Higgs-pairs $(h_u, h_d)$ and in total 75 SM singlets, but no additional (vector-like) exotics. 
Importantly, there are ten flavons $f^0$ that transform as triplets of $\SU{3}_\mathrm{flavor}$ 
(contributing $10 \times 3 = 30$ SM singlets from the point of view of the SM). Their VEVs will be 
important to explain quark and lepton masses and mixings, see section~\ref{sec:AnomalousU1}. 
Concerning hidden sector gaugino condensation, we analyze the chiral spectrum with respect to the 
hidden sector gauge group factor $\SU{5}_\mathrm{hidden}$: The massless spectrum contains two 
$\rep{5}$-plets and two $\crep{5}$-plets. They can decouple without breaking the SM or 
$\SU{5}_\mathrm{hidden}$. Hence, there is no light matter charged under $\SU{5}_\mathrm{hidden}$ 
that enters the beta-function coefficient and we obtain $b=15$.

In total, we have identified 20 almost perfect MSSM-like orbifold models, see the arXiv 
ancillary files~\cite{Parr:2020anc}. All of them are based on the point group $\Z{2}\times\Z{4}$, 
one from the $(1, 6)$ orbifold geometry and 19 from the $(2, 4)$ orbifold geometry. These 20 almost 
perfect MSSM-like orbifold models are very similar from the point of view of the massless matter 
spectrum:
\begin{itemize}
\item there is a gauged $\SU{3}_\mathrm{flavor}$ flavor symmetry,
\item all quark doublets ($q$), up-type quarks ($\bar{u}$), electrons ($\bar{e}$) and up-type 
Higgses ($h_u$) are triplets of $\SU{3}_\mathrm{flavor}$ and live in the bulk of the orbifold,
\item one down-type quark ($\bar{d}$) is a singlet of $\SU{3}_\mathrm{flavor}$ and lives in the bulk of the orbifold,
\item two down-type quarks ($\bar{d}$) and some lepton doublets are localized either in the 
$T_{(1, 3)}$ or in the $T_{(0, 2)}$ twisted sector.
\end{itemize}
We marked these 20 almost perfect MSSM-like orbifold models as yellow points in 
figure~\ref{fig:pred_almost_perfect_MSSM}.

However, it is important to note that a detailed phenomenological study of these models is not 
possible at the moment. The reason is a missing understanding of the $R$-symmetries for these 
orbifold geometries. Even though $R$-symmetries are in general very well understood in heterotic 
orbifold compactifications~\cite{Bizet:2013gf,Nilles:2013lda,Bizet:2013wha}, there are two 
exceptions for orbifolds with $\Z2\times\Z4$ point group, being the $(1, 6)$ and $(2, 4)$ orbifold 
geometries~\cite{Schmitz:2014qoa}. Hence, in order to analyze the phenomenological properties of 
our almost perfect MSSM-like orbifold models in more detail, the $R$-symmetries have to be 
re-analyzed for these orbifolds. 

Finally, note that in our prediction task the $\Z2\times\Z4$ point group is one of the dominant 
classes (with 48,812 MSSM-like orbifold models in total; however, the $(1, 6)$ and $(2, 4)$ 
orbifold geometries of this point group lead to only 82 and 320 MSSM-like orbifold models, 
respectively). Since the almost perfect models are based on the $\Z2\times\Z4$ point group, it is 
reasonable that our ML model predicts $\Z2\times\Z4$ as the orbifold origin of the MSSM for a wide 
range in feature space.

\begin{table}[t!]
\center
\begin{tabular}{|l|r|l|c|}
\hline
sector       & \# & irrep & labels \\
\hline
\hline
$U_1$        &  1 & $\left( \rep{3}; \rep{3},\rep{2}; \rep{1}\right)_{ \tfrac{1}{6}}$ & $q_{i}$ \\
             &  1 & $\left( \rep{1}; \rep{1},\rep{2}; \rep{1}\right)_{-\tfrac{1}{2}}$ & $\ell_{i}$ or $h_d$\\
             &  2 & $\left( \rep{1}; \rep{1},\rep{1}; \rep{1}\right)_0$               & $s_i^0$\\
\hline
$U_2$        &  1 & $\left(\crep{3}; \rep{1},\rep{2}; \rep{1}\right)_{-\tfrac{1}{2}}$ & $\ell_{i}$  or $h_d$\\
             &  1 & $\left( \rep{3}; \rep{1},\rep{2}; \rep{1}\right)_{ \tfrac{1}{2}}$ & ${h_u}$ \\
\hline
$U_3$        &  1 & $\left( \rep{3};\crep{3},\rep{1}; \rep{1}\right)_{-\tfrac{2}{3}}$ & $\bar{u}_{i}$ \\
             &  1 & $\left( \rep{1};\crep{3},\rep{1}; \rep{1}\right)_{ \tfrac{1}{3}}$ & $\bar{d}_{i}$ \\
             &  1 & $\left( \rep{3}; \rep{1},\rep{1}; \rep{1}\right)_1$               & $\bar{e}_{i}$ \\
             &  2 & $\left( \rep{1}; \rep{1},\rep{1};\crep{5}\right)_0$               & $s_i^0$\\
             &  1 & $\left( \rep{1}; \rep{1},\rep{1}; \rep{1}\right)_0$               & $s_i^0$\\
\hline
\hline
$T_{(1, 3)}$ & 22 & $\left( \rep{1}; \rep{1},\rep{1}; \rep{1}\right)_0$               & $s_i^0$\\
             &  2 & $\left( \rep{1};\crep{3},\rep{1}; \rep{1}\right)_{ \tfrac{1}{3}}$ & $\bar{d}_{i}$\\
             &  2 & $\left( \rep{1}; \rep{1},\rep{2}; \rep{1}\right)_{-\tfrac{1}{2}}$ & $\ell_{i}$ or $h_d$\\
             & 10 & $\left(\crep{3}; \rep{1},\rep{1}; \rep{1}\right)_0$               & $f_i^0$\\
             &  2 & $\left( \rep{1}; \rep{1},\rep{1}; \rep{5}\right)_0$               & $s_i^0$\\
\hline
\end{tabular}
\caption{Massless matter spectrum of an ``almost perfect'' MSSM-like orbifold model originating 
from the $\Z{2}\times\Z{4}$ $(2, 4)$ orbifold geometry. The four-dimensional gauge group is 
$\SU{3}_\mathrm{flavor} \times \SU{3}_\mathrm{C} \times \SU{2}_\mathrm{L} \times \SU{5}_\mathrm{hidden} \times \U{1}_\mathrm{Y} \times \U{1}^6$. The untwisted sectors $U_a$, $a=1,2,3$, are defined in section~\ref{sec:HeavyTop}, while 
$T_{(1, 3)}$ denotes the $\theta\,\omega^3$ twisted sector. There are three generations of quarks 
and leptons, three Higgs-pairs but no additional vector-like exotics charged under the SM. 
$\SU{3}_\mathrm{flavor}$ is a gauged flavor symmetry, where $f_i^0$ are so-called flavons: SM 
singlets charged as triplets under $\SU{3}_\mathrm{flavor}$. }
\label{tab:AlmostPerfectMSSM}
\end{table}

%%%%%%%%%%%%%%%%%%%%%%%%%%%%%%%%%%%%%%%%%%%%%%%%%%%%%%%%%%%%%%%%%%%%%%%%%%%%%%%%%%%%%%%%%%%%%%%%%%%%%%%%%%%%%%%%%%%%%%%%%%%%%%%%%%%%%%%%%%
\section{Conclusions and outlook}
\label{sec:conclusion}

In conclusion, in this paper we show that the huge number of possible string models in the 
heterotic orbifold landscape gets subdivided into several sub-landscapes, each with their own 
distinct phenomenological properties. These properties include most prominently the appearance of 
various types of vector-like exotics (see tables~\ref{tab:MSSMsummary1} and~\ref{tab:MSSMsummary2}), 
the number of SM singlets, the existence of non-Abelian flavor symmetries, and the hidden sector 
beta function relevant for supersymmetry breaking via gaugino condensation, among others.

In detail, we demonstrate that the choice of the point group $P$ that underlies an orbifold geometry 
leaves a particular imprint, for example, on the particle spectrum of vector-like exotics for 
MSSM-like orbifold models. This imprint can be exploited in order to predict with high certainty 
the most likely point group that is able to reproduce a specific MSSM-like particle spectrum, see 
table~\ref{tab:classification_report}. In more detail, this is achieved using a machine learning 
(ML) algorithm known as ``boosted decision tree'' that is particularly suitable for our 
classification task. We train and test our boosted decision tree on the largest known set 
of 126,783 distinct MSSM-like orbifold models. Then, we dissect our trained ensemble of decision 
trees in order to identify the most important phenomenological properties used by the decision 
trees to classify the point group of a given MSSM-like particle spectrum, see 
figure~\ref{fig:FeaturesInDecisionTree}.

After training and evaluating our ML model, we apply our boosted decision tree to the MSSM in order 
to predict the stringy origin of the MSSM. For this task, we assume that supersymmetry is broken in 
the MSSM by hidden sector gaugino condensation, and we extend the MSSM by a large number of SM 
singlets (i.e.\ right-handed neutrinos). Then, we vary over (i) the beta-function coefficient $b$ 
of the hidden sector gauge group and (ii) the number of SM singlets, and predict in each case the 
most probable stringy origin of the MSSM. The result is shown in figure~\ref{fig:pred_almost_perfect_MSSM}. 
In a nutshell, we find that for up to $\approx 140$ SM singlets and all ranges of $b$, orbifolds 
with $\Z{2}\times\Z{4}$ point group seem to be the most promising orbifold compactifications of the 
heterotic string to yield the MSSM. For an even larger number of SM singlets, the boosted decision tree 
predicts orbifolds with $\Z{4}\times\Z{4}$ point group. As one can see in figure~\ref{fig:pred_MSSM} 
these predictions depend on a few additional phenomenological features that are inspired by the 
MSSM-like orbifold models of our dataset. Varying these features (in addition to $b$ and the number 
of SM singlets) yields further promising point groups, like $\Z{2}\times\Z{2}$ and in small corners 
of the parameter space also $\Z{6}$-II, $\Z{8}$-I and $\Z{3}\times\Z{6}$. Hence, we suggest to 
focus heterotic orbifold model building on one of these most promising orbifold geometries.

Furthermore, we present the first ``almost perfect'' MSSM-like orbifold models, see 
table~\ref{tab:AlmostPerfectMSSM} for one example. These orbifold models, which were unknown in the 
literature, have exactly three generations of quarks and leptons, either three or five Higgs-pairs, 
but no additional vector-like exotics. They originate from the $\Z{2}\times\Z{4}$ $(1,6)$ and 
$(2,4)$ orbifold geometries. These orbifolds are equipped with non-local GUT 
breaking~\cite{Fischer:2012qj}, relevant for stringy dark matter~\cite{Mutter:2019byx}. However, 
since $R$-symmetries are not under control just for these two $\Z{2}\times\Z{4}$ orbifold 
geometries~\cite{Schmitz:2014qoa}, our findings should encourage to continue the efforts of 
refs.~\cite{Bizet:2013gf,Nilles:2013lda,Bizet:2013wha} to study the $R$-symmetries for these 
special orbifold geometries.

\section*{Acknowledgments}

This work is supported by the Deutsche Forschungsgemeinschaft (SFB1258). We would like to thank 
James Halverson, Sven Krippendorf and Sa{\'u}l Ramos-S{\'a}nchez for useful discussions.

\clearpage

%%%%%%%%%%%%%%%%%%%%%%%%%%%%%%%%%%%%%%%%%%%%%%%%%%%%%%%%%%%%%%%%%%%%%%%%%%%%%%%%%%%%%%%%%%%%%%%%%%%%%%%%%%%%%%%%%%%%%%%%%%%%%%%%%%%%%%%%%%
\appendix

\vspace*{-3.3cm}
\section[Vector-like exotics in the Orbifold-Landscape]{Vector-like exotics in the Orbifold-Landscape}
\label{app:exotics}
\vspace*{-0.5cm}

\begin{table}[H]
{\notsotiny
\Rotatebox{90}{%
\begin{tabular}{|l|r|r|r|r|r|r|r|r|r|r|r|r|r|r|r|}
\hline
          & $\Z{4}$ & $\Z{6}$-I & $\Z{6}$-II & $\Z{7}$ & $\Z{8}$-I & $\Z{8}$-II & $\Z{12}$-I & $\Z{12}$-II & $\Z{2}\times\Z{2}$ & $\Z{2}\times\Z{4}$ & $\Z{2}\times\Z{6}$-I & $\Z{3}\times\Z{3}$ & $\Z{3}\times\Z{6}$ & $\Z{4}\times\Z{4}$ & $\Z{6}\times\Z{6}$ \\
\hline
\# MSSM   & $212$   & $62$      & $1,870$    & $1$     & $865$     & $2,844$    & $1,250$    & $435$       & $1,711$            & $55,429$           & $1,095$              & $3,337$            & $5,153$            & $48,812$           & $3,707$\\
\hline
\hline
$( \rep{3}, \rep{2})_{\nicefrac{1}{6}}$ 	& $1.89 \%$ 	& $0 \%$ 	& $31.60 \%$ 	& $0 \%$ 	& $4.05 \%$ 	& $25.00 \%$ 	& $4.00 \%$ 	& $24.14 \%$ 	& $8.53 \%$ 	& $14.04 \%$ 	& $31.60 \%$ 	& $14.59 \%$ 	& $19.04 \%$ 	& $15.31 \%$ 	& $39.60 \%$ 	\\
$(\crep{3}, \rep{1})_{\nicefrac{-2}{3}}$ 	& $50.47 \%$ 	& $0 \%$ 	& $28.66 \%$ 	& $0 \%$ 	& $5.55 \%$ 	& $44.13 \%$ 	& $3.28 \%$ 	& $35.63 \%$ 	& $8.36 \%$ 	& $14.35 \%$ 	& $22.92 \%$ 	& $18.91 \%$ 	& $18.84 \%$ 	& $14.93 \%$ 	& $29.92 \%$ 	\\
$(\crep{3}, \rep{1})_{\nicefrac{1}{3}}$ 	& $100 \%$ 	& $100 \%$ 	& $99.95 \%$ 	& $100 \%$ 	& $99.54 \%$ 	& $100 \%$ 	& $100 \%$ 	& $100 \%$ 	& $100 \%$ 	& $98.19 \%$ 	& $99.73 \%$ 	& $94.76 \%$ 	& $99.83 \%$ 	& $99.03 \%$ 	& $99.00 \%$ 	\\
$( \rep{1}, \rep{2})_{\nicefrac{-1}{2}}$ 	& $96.23 \%$ 	& $35.48 \%$ 	& $92.19 \%$ 	& $100 \%$ 	& $93.99 \%$ 	& $94.94 \%$ 	& $78.96 \%$ 	& $91.72 \%$ 	& $92.17 \%$ 	& $92.20 \%$ 	& $96.26 \%$ 	& $90.62 \%$ 	& $99.20 \%$ 	& $95.07 \%$ 	& $95.50 \%$ 	\\
$( \rep{1}, \rep{1})_1$ 	& $1.89 \%$ 	& $0 \%$ 	& $28.66 \%$ 	& $0 \%$ 	& $5.55 \%$ 	& $44.09 \%$ 	& $3.28 \%$ 	& $35.63 \%$ 	& $8.36 \%$ 	& $14.36 \%$ 	& $22.92 \%$ 	& $11.30 \%$ 	& $18.92 \%$ 	& $14.98 \%$ 	& $29.92 \%$ 	\\
\hline
$( \rep{1}, \rep{1})_0$ 	& $100 \%$ 	& $100 \%$ 	& $100 \%$ 	& $100 \%$ 	& $100 \%$ 	& $100 \%$ 	& $100 \%$ 	& $100 \%$ 	& $100 \%$ 	& $100 \%$ 	& $100 \%$ 	& $100 \%$ 	& $100 \%$ 	& $100 \%$ 	& $100 \%$ 	\\
\hline
$( \rep{3}, \rep{2})_{\nicefrac{-1}{2}}$ 	& $0 \%$ 	& $0 \%$ 	& $0 \%$ 	& $0 \%$ 	& $0 \%$ 	& $0 \%$ 	& $0 \%$ 	& $0 \%$ 	& $0 \%$ 	& $0 \%$ 	& $0 \%$ 	& $0.06 \%$ 	& $0 \%$ 	& $0 \%$ 	& $0 \%$ 	\\
$( \rep{3}, \rep{2})_{\nicefrac{1}{2}}$ 	& $0 \%$ 	& $0 \%$ 	& $0 \%$ 	& $0 \%$ 	& $0 \%$ 	& $0 \%$ 	& $0 \%$ 	& $0 \%$ 	& $0 \%$ 	& $0 \%$ 	& $0 \%$ 	& $0.18 \%$ 	& $0 \%$ 	& $0 \%$ 	& $0 \%$ 	\\
$( \rep{3}, \rep{2})_{\nicefrac{-1}{3}}$ 	& $0 \%$ 	& $0 \%$ 	& $0 \%$ 	& $0 \%$ 	& $0.23 \%$ 	& $1.05 \%$ 	& $0.24 \%$ 	& $0 \%$ 	& $0 \%$ 	& $2.27 \%$ 	& $0.09 \%$ 	& $0 \%$ 	& $0.43 \%$ 	& $0.53 \%$ 	& $0.32 \%$ 	\\
$( \rep{3}, \rep{2})_{\nicefrac{-1}{6}}$ 	& $0 \%$ 	& $0 \%$ 	& $0 \%$ 	& $0 \%$ 	& $0 \%$ 	& $0 \%$ 	& $0.64 \%$ 	& $0 \%$ 	& $0 \%$ 	& $0 \%$ 	& $0 \%$ 	& $7.19 \%$ 	& $0.02 \%$ 	& $0 \%$ 	& $0 \%$ 	\\
$( \rep{3}, \rep{2})_{\nicefrac{-1}{12}}$ 	& $2.83 \%$ 	& $0 \%$ 	& $0 \%$ 	& $0 \%$ 	& $0 \%$ 	& $0 \%$ 	& $0 \%$ 	& $0 \%$ 	& $0 \%$ 	& $0.00 \%$ 	& $0 \%$ 	& $0 \%$ 	& $0 \%$ 	& $0.00 \%$ 	& $0 \%$ 	\\
\hline
$( \rep{3}, \rep{1})_0$ 	& $0 \%$ 	& $0 \%$ 	& $6.84 \%$ 	& $0 \%$ 	& $0 \%$ 	& $0 \%$ 	& $25.12 \%$ 	& $0 \%$ 	& $0 \%$ 	& $0 \%$ 	& $0 \%$ 	& $85.77 \%$ 	& $9.99 \%$ 	& $0 \%$ 	& $0.35 \%$ 	\\
$( \rep{3}, \rep{1})_{\nicefrac{-1}{2}}$ 	& $0 \%$ 	& $0 \%$ 	& $0 \%$ 	& $0 \%$ 	& $0 \%$ 	& $0 \%$ 	& $1.12 \%$ 	& $0 \%$ 	& $0 \%$ 	& $0 \%$ 	& $0 \%$ 	& $0 \%$ 	& $0.06 \%$ 	& $0 \%$ 	& $0 \%$ 	\\
$( \rep{3}, \rep{1})_{\nicefrac{1}{2}}$ 	& $0 \%$ 	& $0 \%$ 	& $0 \%$ 	& $0 \%$ 	& $0 \%$ 	& $0 \%$ 	& $0.80 \%$ 	& $0 \%$ 	& $0 \%$ 	& $0 \%$ 	& $0 \%$ 	& $0 \%$ 	& $0.04 \%$ 	& $0 \%$ 	& $0 \%$ 	\\
$( \rep{3}, \rep{1})_{\nicefrac{-2}{3}}$ 	& $0 \%$ 	& $0 \%$ 	& $0 \%$ 	& $0 \%$ 	& $0 \%$ 	& $0 \%$ 	& $0.32 \%$ 	& $0 \%$ 	& $0 \%$ 	& $0 \%$ 	& $0 \%$ 	& $5.60 \%$ 	& $0.06 \%$ 	& $0 \%$ 	& $0 \%$ 	\\
$( \rep{3}, \rep{1})_{\nicefrac{1}{3}}$ 	& $0 \%$ 	& $0 \%$ 	& $4.71 \%$ 	& $0 \%$ 	& $0 \%$ 	& $0 \%$ 	& $18.72 \%$ 	& $0 \%$ 	& $0 \%$ 	& $0 \%$ 	& $0 \%$ 	& $73.24 \%$ 	& $6.85 \%$ 	& $0 \%$ 	& $0.27 \%$ 	\\
$( \rep{3}, \rep{1})_{\nicefrac{-5}{6}}$ 	& $0 \%$ 	& $0 \%$ 	& $2.89 \%$ 	& $0 \%$ 	& $0 \%$ 	& $0.74 \%$ 	& $0 \%$ 	& $0 \%$ 	& $0 \%$ 	& $0.12 \%$ 	& $0 \%$ 	& $0 \%$ 	& $0.62 \%$ 	& $0.44 \%$ 	& $0.76 \%$ 	\\
$( \rep{3}, \rep{1})_{\nicefrac{-1}{6}}$ 	& $0 \%$ 	& $0 \%$ 	& $0.21 \%$ 	& $0 \%$ 	& $0 \%$ 	& $0 \%$ 	& $4.32 \%$ 	& $0 \%$ 	& $0 \%$ 	& $0 \%$ 	& $0 \%$ 	& $0 \%$ 	& $0.45 \%$ 	& $0 \%$ 	& $0 \%$ 	\\
$( \rep{3}, \rep{1})_{\nicefrac{1}{6}}$ 	& $67.45 \%$ 	& $93.55 \%$ 	& $66.47 \%$ 	& $0 \%$ 	& $87.63 \%$ 	& $74.30 \%$ 	& $69.20 \%$ 	& $66.90 \%$ 	& $85.21 \%$ 	& $90.85 \%$ 	& $86.58 \%$ 	& $0 \%$ 	& $84.84 \%$ 	& $95.75 \%$ 	& $98.17 \%$ 	\\
$( \rep{3}, \rep{1})_{\nicefrac{-7}{12}}$ 	& $2.83 \%$ 	& $0 \%$ 	& $0 \%$ 	& $0 \%$ 	& $0 \%$ 	& $0 \%$ 	& $0.32 \%$ 	& $0 \%$ 	& $0 \%$ 	& $0 \%$ 	& $0 \%$ 	& $0 \%$ 	& $0 \%$ 	& $0.01 \%$ 	& $0 \%$ 	\\
$( \rep{3}, \rep{1})_{\nicefrac{-1}{12}}$ 	& $41.04 \%$ 	& $0 \%$ 	& $0 \%$ 	& $0 \%$ 	& $0.23 \%$ 	& $0.56 \%$ 	& $1.76 \%$ 	& $0 \%$ 	& $0 \%$ 	& $0.08 \%$ 	& $0 \%$ 	& $0 \%$ 	& $0 \%$ 	& $0.13 \%$ 	& $0 \%$ 	\\
$( \rep{3}, \rep{1})_{\nicefrac{5}{12}}$ 	& $4.72 \%$ 	& $0 \%$ 	& $0 \%$ 	& $0 \%$ 	& $0 \%$ 	& $0.14 \%$ 	& $0.80 \%$ 	& $0 \%$ 	& $0 \%$ 	& $0.03 \%$ 	& $0 \%$ 	& $0 \%$ 	& $0 \%$ 	& $0.06 \%$ 	& $0 \%$ 	\\
$( \rep{3}, \rep{1})_{\nicefrac{2}{21}}$ 	& $0 \%$ 	& $0 \%$ 	& $0 \%$ 	& $100 \%$ 	& $0 \%$ 	& $0 \%$ 	& $0 \%$ 	& $0 \%$ 	& $0 \%$ 	& $0 \%$ 	& $0 \%$ 	& $0 \%$ 	& $0 \%$ 	& $0 \%$ 	& $0 \%$ 	\\
$( \rep{3}, \rep{1})_{\nicefrac{5}{21}}$ 	& $0 \%$ 	& $0 \%$ 	& $0 \%$ 	& $100 \%$ 	& $0 \%$ 	& $0 \%$ 	& $0 \%$ 	& $0 \%$ 	& $0 \%$ 	& $0 \%$ 	& $0 \%$ 	& $0 \%$ 	& $0 \%$ 	& $0 \%$ 	& $0 \%$ 	\\
\hline
$( \rep{1}, \rep{2})_0$ 	& $82.55 \%$ 	& $100 \%$ 	& $85.94 \%$ 	& $0 \%$ 	& $99.31 \%$ 	& $93.95 \%$ 	& $76.32 \%$ 	& $90.34 \%$ 	& $84.57 \%$ 	& $96.49 \%$ 	& $98.17 \%$ 	& $0 \%$ 	& $89.93 \%$ 	& $99.26 \%$ 	& $99.57 \%$ 	\\
$( \rep{1}, \rep{2})_{\nicefrac{1}{3}}$ 	& $0 \%$ 	& $0 \%$ 	& $0.43 \%$ 	& $0 \%$ 	& $0 \%$ 	& $0 \%$ 	& $4.72 \%$ 	& $0 \%$ 	& $0 \%$ 	& $0 \%$ 	& $0 \%$ 	& $0 \%$ 	& $0.66 \%$ 	& $0 \%$ 	& $0 \%$ 	\\
$( \rep{1}, \rep{2})_{\nicefrac{2}{3}}$ 	& $0 \%$ 	& $0 \%$ 	& $0 \%$ 	& $0 \%$ 	& $0 \%$ 	& $0 \%$ 	& $0.80 \%$ 	& $0 \%$ 	& $0 \%$ 	& $0 \%$ 	& $0 \%$ 	& $0 \%$ 	& $0.02 \%$ 	& $0 \%$ 	& $0 \%$ 	\\
$( \rep{1}, \rep{2})_{\nicefrac{1}{4}}$ 	& $41.98 \%$ 	& $0 \%$ 	& $0 \%$ 	& $0 \%$ 	& $0.23 \%$ 	& $0.70 \%$ 	& $2.40 \%$ 	& $0 \%$ 	& $0 \%$ 	& $0.08 \%$ 	& $0 \%$ 	& $0 \%$ 	& $0 \%$ 	& $0.14 \%$ 	& $0 \%$ 	\\
$( \rep{1}, \rep{2})_{\nicefrac{3}{4}}$ 	& $2.83 \%$ 	& $0 \%$ 	& $0 \%$ 	& $0 \%$ 	& $0 \%$ 	& $0 \%$ 	& $0 \%$ 	& $0 \%$ 	& $0 \%$ 	& $0 \%$ 	& $0 \%$ 	& $0 \%$ 	& $0 \%$ 	& $0 \%$ 	& $0 \%$ 	\\
$( \rep{1}, \rep{2})_{\nicefrac{1}{6}}$ 	& $0 \%$ 	& $0 \%$ 	& $8.98 \%$ 	& $0 \%$ 	& $0 \%$ 	& $0 \%$ 	& $26.72 \%$ 	& $0 \%$ 	& $0 \%$ 	& $0 \%$ 	& $0 \%$ 	& $96.55 \%$ 	& $10.58 \%$ 	& $0 \%$ 	& $0.35 \%$ 	\\
$( \rep{1}, \rep{2})_{\nicefrac{5}{6}}$ 	& $0 \%$ 	& $0 \%$ 	& $0 \%$ 	& $0 \%$ 	& $0 \%$ 	& $0 \%$ 	& $0 \%$ 	& $0 \%$ 	& $0 \%$ 	& $0 \%$ 	& $0 \%$ 	& $0.21 \%$ 	& $0 \%$ 	& $0 \%$ 	& $0 \%$ 	\\
$( \rep{1}, \rep{2})_{\nicefrac{1}{14}}$ 	& $0 \%$ 	& $0 \%$ 	& $0 \%$ 	& $100 \%$ 	& $0 \%$ 	& $0 \%$ 	& $0 \%$ 	& $0 \%$ 	& $0 \%$ 	& $0 \%$ 	& $0 \%$ 	& $0 \%$ 	& $0 \%$ 	& $0 \%$ 	& $0 \%$ 	\\
$( \rep{1}, \rep{2})_{\nicefrac{5}{14}}$ 	& $0 \%$ 	& $0 \%$ 	& $0 \%$ 	& $100 \%$ 	& $0 \%$ 	& $0 \%$ 	& $0 \%$ 	& $0 \%$ 	& $0 \%$ 	& $0 \%$ 	& $0 \%$ 	& $0 \%$ 	& $0 \%$ 	& $0 \%$ 	& $0 \%$ 	\\
\hline
$( \rep{1}, \rep{1})_{\nicefrac{1}{2}}$ 	& $100 \%$ 	& $100 \%$ 	& $92.09 \%$ 	& $0 \%$ 	& $100 \%$ 	& $100 \%$ 	& $76.96 \%$ 	& $100 \%$ 	& $95.79 \%$ 	& $99.16 \%$ 	& $99.36 \%$ 	& $0 \%$ 	& $90.28 \%$ 	& $99.99 \%$ 	& $99.65 \%$ 	\\
$( \rep{1}, \rep{1})_{\nicefrac{1}{3}}$ 	& $0 \%$ 	& $0 \%$ 	& $8.98 \%$ 	& $0 \%$ 	& $0 \%$ 	& $0 \%$ 	& $29.20 \%$ 	& $0 \%$ 	& $0 \%$ 	& $0 \%$ 	& $0 \%$ 	& $99.52 \%$ 	& $10.58 \%$ 	& $0 \%$ 	& $0.35 \%$ 	\\
$( \rep{1}, \rep{1})_{\nicefrac{2}{3}}$ 	& $0 \%$ 	& $0 \%$ 	& $8.98 \%$ 	& $0 \%$ 	& $0 \%$ 	& $0 \%$ 	& $28.72 \%$ 	& $0 \%$ 	& $0 \%$ 	& $0 \%$ 	& $0 \%$ 	& $95.62 \%$ 	& $10.58 \%$ 	& $0 \%$ 	& $0.35 \%$ 	\\
$( \rep{1}, \rep{1})_{\nicefrac{1}{4}}$ 	& $52.36 \%$ 	& $0 \%$ 	& $0 \%$ 	& $0 \%$ 	& $0.23 \%$ 	& $0.95 \%$ 	& $3.20 \%$ 	& $0 \%$ 	& $0 \%$ 	& $0.19 \%$ 	& $0 \%$ 	& $0 \%$ 	& $0 \%$ 	& $0.17 \%$ 	& $0 \%$ 	\\
$( \rep{1}, \rep{1})_{\nicefrac{3}{4}}$ 	& $42.92 \%$ 	& $0 \%$ 	& $0 \%$ 	& $0 \%$ 	& $0.23 \%$ 	& $0.42 \%$ 	& $1.92 \%$ 	& $0 \%$ 	& $0 \%$ 	& $0.12 \%$ 	& $0 \%$ 	& $0 \%$ 	& $0 \%$ 	& $0.14 \%$ 	& $0 \%$ 	\\
$( \rep{1}, \rep{1})_{\nicefrac{1}{6}}$ 	& $0 \%$ 	& $0 \%$ 	& $1.07 \%$ 	& $0 \%$ 	& $0 \%$ 	& $0 \%$ 	& $6.32 \%$ 	& $0 \%$ 	& $0 \%$ 	& $0 \%$ 	& $0 \%$ 	& $0 \%$ 	& $0.85 \%$ 	& $0 \%$ 	& $0 \%$ 	\\
$( \rep{1}, \rep{1})_{\nicefrac{5}{6}}$ 	& $0 \%$ 	& $0 \%$ 	& $0 \%$ 	& $0 \%$ 	& $0 \%$ 	& $0 \%$ 	& $1.12 \%$ 	& $0 \%$ 	& $0 \%$ 	& $0 \%$ 	& $0 \%$ 	& $0 \%$ 	& $0.04 \%$ 	& $0 \%$ 	& $0 \%$ 	\\
$( \rep{1}, \rep{1})_{\nicefrac{1}{7}}$ 	& $0 \%$ 	& $0 \%$ 	& $0 \%$ 	& $100 \%$ 	& $0 \%$ 	& $0 \%$ 	& $0 \%$ 	& $0 \%$ 	& $0 \%$ 	& $0 \%$ 	& $0 \%$ 	& $0 \%$ 	& $0 \%$ 	& $0 \%$ 	& $0 \%$ 	\\
$( \rep{1}, \rep{1})_{\nicefrac{2}{7}}$ 	& $0 \%$ 	& $0 \%$ 	& $0 \%$ 	& $100 \%$ 	& $0 \%$ 	& $0 \%$ 	& $0 \%$ 	& $0 \%$ 	& $0 \%$ 	& $0 \%$ 	& $0 \%$ 	& $0 \%$ 	& $0 \%$ 	& $0 \%$ 	& $0 \%$ 	\\
$( \rep{1}, \rep{1})_{\nicefrac{3}{7}}$ 	& $0 \%$ 	& $0 \%$ 	& $0 \%$ 	& $100 \%$ 	& $0 \%$ 	& $0 \%$ 	& $0 \%$ 	& $0 \%$ 	& $0 \%$ 	& $0 \%$ 	& $0 \%$ 	& $0 \%$ 	& $0 \%$ 	& $0 \%$ 	& $0 \%$ 	\\
$( \rep{1}, \rep{1})_{\nicefrac{4}{7}}$ 	& $0 \%$ 	& $0 \%$ 	& $0 \%$ 	& $100 \%$ 	& $0 \%$ 	& $0 \%$ 	& $0 \%$ 	& $0 \%$ 	& $0 \%$ 	& $0 \%$ 	& $0 \%$ 	& $0 \%$ 	& $0 \%$ 	& $0 \%$ 	& $0 \%$ 	\\
\hline
\end{tabular}
}}%
\caption{Percentage of MSSM-like orbifold models with certain types of vector-like exotics. 
Hypercharge is normalized such that $( \rep{3}, \rep{2})_{\nicefrac{1}{6}}$ is a left-chiral 
quark-doublet. The row ``\# MSSM'' lists the number of inequivalent MSSM-like orbifold models in 
our dataset. A complex representation has to be amended by its complex conjugate, e.g.\ 
$( \rep{3}, \rep{2})_{\nicefrac{1}{6}}$ stands for $( \rep{3}, \rep{2})_{\nicefrac{1}{6}} \oplus (\crep{3}, \rep{2})_{\nicefrac{-1}{6}}$.}
\label{tab:MSSMsummary1}
\end{table}

\begin{table}[H]
\Rotatebox{90}{%
{\notsotiny
\begin{tabular}{|l|r|r|r|r|r|r|r|r|r|r|r|r|r|r|r|}
\hline
          & $\Z{4}$ & $\Z{6}$-I & $\Z{6}$-II & $\Z{7}$ & $\Z{8}$-I & $\Z{8}$-II & $\Z{12}$-I & $\Z{12}$-II & $\Z{2}\times\Z{2}$ & $\Z{2}\times\Z{4}$ & $\Z{2}\times\Z{6}$-I & $\Z{3}\times\Z{3}$ & $\Z{3}\times\Z{6}$ & $\Z{4}\times\Z{4}$ & $\Z{6}\times\Z{6}$ \\
\hline
\# MSSM   & $212$   & $62$      & $1,870$    & $1$     & $865$     & $2,844$    & $1,250$    & $435$       & $1,711$            & $55,429$           & $1,095$              & $3,337$            & $5,153$            & $48,812$           & $3,707$\\
\hline
\hline
$( \rep{3}, \rep{2})_{\nicefrac{1}{6}}$ 	& $0.04$ 	& $0$ 	& $0.48$ 	& $0$ 	& $0.05$ 	& $0.31$ 	& $0.09$ 	& $0.46$ 	& $0.09$ 	& $0.19$ 	& $0.51$ 	& $0.17$ 	& $0.23$ 	& $0.18$ 	& $0.46$ 	\\
$(\crep{3}, \rep{1})_{\nicefrac{-2}{3}}$ 	& $0.52$ 	& $0$ 	& $0.41$ 	& $0$ 	& $0.08$ 	& $0.49$ 	& $0.06$ 	& $0.71$ 	& $0.09$ 	& $0.21$ 	& $0.35$ 	& $0.23$ 	& $0.23$ 	& $0.18$ 	& $0.36$ 	\\
$(\crep{3}, \rep{1})_{\nicefrac{1}{3}}$ 	& $3.52$ 	& $3.52$ 	& $5.22$ 	& $4.00$ 	& $4.41$ 	& $4.36$ 	& $4.21$ 	& $7.58$ 	& $7.31$ 	& $5.66$ 	& $7.96$ 	& $3.64$ 	& $6.57$ 	& $5.72$ 	& $6.25$ 	\\
$( \rep{1}, \rep{2})_{\nicefrac{-1}{2}}$ 	& $2.64$ 	& $1.68$ 	& $4.03$ 	& $3.00$ 	& $3.41$ 	& $3.06$ 	& $2.92$ 	& $4.28$ 	& $3.34$ 	& $3.97$ 	& $5.23$ 	& $2.86$ 	& $5.83$ 	& $4.46$ 	& $5.50$ 	\\
$( \rep{1}, \rep{1})_1$ 	& $0.04$ 	& $0$ 	& $0.41$ 	& $0$ 	& $0.08$ 	& $0.49$ 	& $0.06$ 	& $0.71$ 	& $0.09$ 	& $0.21$ 	& $0.35$ 	& $0.15$ 	& $0.23$ 	& $0.18$ 	& $0.36$ 	\\
\hline
$( \rep{1}, \rep{1})_0$ 	& $67.37$ 	& $136.58$ 	& $106.06$ 	& $32.00$ 	& $108.74$ 	& $89.78$ 	& $85.99$ 	& $101.78$ 	& $122.54$ 	& $132.64$ 	& $139.59$ 	& $106.48$ 	& $177.56$ 	& $166.38$ 	& $186.95$ 	\\
\hline
$( \rep{3}, \rep{2})_{\nicefrac{-1}{2}}$ 	& $0$ 	& $0$ 	& $0$ 	& $0$ 	& $0$ 	& $0$ 	& $0$ 	& $0$ 	& $0$   & $0$ 	& $0$ 	& $0.00$& $0$ 	& $0$ 	& $0$ 	\\
$( \rep{3}, \rep{2})_{\nicefrac{1}{2}}$ 	& $0$ 	& $0$ 	& $0$ 	& $0$ 	& $0$ 	& $0$ 	& $0$ 	& $0$ 	& $0$   & $0$ 	& $0$ 	& $0.00$& $0$ 	& $0$ 	& $0$ 	\\
$( \rep{3}, \rep{2})_{\nicefrac{-1}{3}}$ 	& $0$ 	& $0$ 	& $0$ 	& $0$ 	& $0.00$& $0.01$& $0.00$& $0$ 	& $0$ 	& $0.03$& $0.00$& $0$ 	& $0.01$& $0.01$& $0.00$\\
$( \rep{3}, \rep{2})_{\nicefrac{-1}{6}}$ 	& $0$ 	& $0$ 	& $0$ 	& $0$ 	& $0$ 	& $0$ 	& $0.01$& $0$ 	& $0$ 	& $0$ 	& $0$ 	& $0.08$& $0.00$& $0$ 	& $0$ 	\\
$( \rep{3}, \rep{2})_{\nicefrac{-1}{12}}$ 	& $0.03$& $0$ 	& $0$ 	& $0$ 	& $0$ 	& $0$ 	& $0$ 	& $0$ 	& $0$ 	& $0.00$& $0$ 	& $0$ 	& $0$ 	& $0.00$& $0$ 	\\
\hline
$( \rep{3}, \rep{1})_0$ 	                & $0$ 	& $0$ 	& $0.19$& $0$ 	& $0$ 	& $0$ 	& $0.69$& $0$ 	& $0$ 	& $0$ 	& $0$ 	& $3.32$& $0.36$& $0$ 	& $0.01$\\
$( \rep{3}, \rep{1})_{\nicefrac{-1}{2}}$ 	& $0$ 	& $0$ 	& $0$ 	& $0$ 	& $0$ 	& $0$ 	& $0.01$& $0$ 	& $0$ 	& $0$ 	& $0$ 	& $0$ 	& $0.00$& $0$ 	& $0$ 	\\
$( \rep{3}, \rep{1})_{\nicefrac{1}{2}}$ 	& $0$ 	& $0$ 	& $0$ 	& $0$ 	& $0$ 	& $0$ 	& $0.01$& $0$ 	& $0$ 	& $0$ 	& $0$ 	& $0$ 	& $0.00$& $0$ 	& $0$ 	\\
$( \rep{3}, \rep{1})_{\nicefrac{-2}{3}}$ 	& $0$ 	& $0$ 	& $0$ 	& $0$ 	& $0$ 	& $0$ 	& $0.00$& $0$ 	& $0$ 	& $0$ 	& $0$ 	& $0.06$& $0.00$& $0$ 	& $0$ 	\\
$( \rep{3}, \rep{1})_{\nicefrac{1}{3}}$ 	& $0$ 	& $0$ 	& $0.13$& $0$ 	& $0$ 	& $0$ 	& $0.36$& $0$ 	& $0$ 	& $0$ 	& $0$ 	& $1.95$& $0.14$& $0$ 	& $0.01$\\
$( \rep{3}, \rep{1})_{\nicefrac{-5}{6}}$ 	& $0$ 	& $0$ 	& $0.04$& $0$ 	& $0$ 	& $0.01$& $0$ 	& $0$ 	& $0$ 	& $0.00$& $0$ 	& $0$ 	& $0.01$& $0.01$& $0.01$\\
$( \rep{3}, \rep{1})_{\nicefrac{-1}{6}}$ 	& $0$ 	& $0$ 	& $0.00$& $0$ 	& $0$ 	& $0$ 	& $0.06$& $0$ 	& $0$ 	& $0$ 	& $0$ 	& $0$ 	& $0.01$& $0$ 	& $0$ 	\\
$( \rep{3}, \rep{1})_{\nicefrac{1}{6}}$ 	& $1.99$& $3.29$& $2.38$& $0$ 	& $3.14$& $2.55$& $2.00$& $1.69$& $3.83$& $4.59$& $4.87$& $0$ 	& $3.47$& $5.37$& $6.57$\\
$( \rep{3}, \rep{1})_{\nicefrac{-7}{12}}$ 	& $0.03$& $0$ 	& $0$ 	& $0$ 	& $0$ 	& $0$ 	& $0.00$& $0$ 	& $0$ 	& $0$ 	& $0$ 	& $0$ 	& $0$ 	& $0.00$& $0$ 	\\
$( \rep{3}, \rep{1})_{\nicefrac{-1}{12}}$ 	& $1.10$& $0$ 	& $0$ 	& $0$ 	& $0.00$& $0.01$& $0.04$& $0$ 	& $0$ 	& $0.00$& $0$ 	& $0$ 	& $0$ 	& $0.00$& $0$ 	\\
$( \rep{3}, \rep{1})_{\nicefrac{5}{12}}$ 	& $0.08$& $0$ 	& $0$ 	& $0$ 	& $0$ 	& $0.00$& $0.01$& $0$ 	& $0$ 	& $0.00$& $0$ 	& $0$ 	& $0$ 	& $0.00$& $0$ 	\\
$( \rep{3}, \rep{1})_{\nicefrac{2}{21}}$ 	& $0$ 	& $0$ 	& $0$ 	& $3.00$& $0$ 	& $0$ 	& $0$ 	& $0$ 	& $0$ 	& $0$ 	& $0$ 	& $0$ 	& $0$ 	& $0$ 	& $0$ 	\\
$( \rep{3}, \rep{1})_{\nicefrac{5}{21}}$ 	& $0$ 	& $0$ 	& $0$ 	& $4.00$& $0$ 	& $0$ 	& $0$ 	& $0$ 	& $0$   & $0$ 	& $0$ 	& $0$ 	& $0$ 	& $0$ 	& $0$ 	\\
\hline
$( \rep{1}, \rep{2})_0$ 	& $7.78$ 	& $8.58$& $7.99$ 	& $0$ 	& $9.49$& $8.23$& $5.84$& $6.10$& $7.62$&$13.58$&$14.25$& $0$ 	&$10.97$&$15.88$&$19.27$\\
$( \rep{1}, \rep{2})_{\nicefrac{1}{3}}$ 	& $0$ 	& $0$ 	& $0.01$& $0$ 	& $0$ 	& $0$ 	& $0.08$& $0$ 	& $0$	& $0$ 	& $0$ 	& $0$ 	& $0.01$& $0$ 	& $0$ 	\\
$( \rep{1}, \rep{2})_{\nicefrac{2}{3}}$ 	& $0$ 	& $0$ 	& $0$ 	& $0$ 	& $0$ 	& $0$ 	& $0.01$& $0$ 	& $0$	& $0$ 	& $0$ 	& $0$ 	& $0.00$& $0$ 	& $0$ 	\\
$( \rep{1}, \rep{2})_{\nicefrac{1}{4}}$ 	& $1.19$& $0$ 	& $0$ 	& $0$ 	& $0.00$& $0.02$& $0.07$& $0$ 	& $0$	& $0.00$& $0$ 	& $0$ 	& $0$ 	& $0.01$& $0$ 	\\
$( \rep{1}, \rep{2})_{\nicefrac{3}{4}}$ 	& $0.03$& $0$ 	& $0$ 	& $0$ 	& $0$ 	& $0$ 	& $0$ 	& $0$ 	& $0$	& $0$ 	& $0$ 	& $0$ 	& $0$ 	& $0$ 	& $0$ 	\\
$( \rep{1}, \rep{2})_{\nicefrac{1}{6}}$ 	& $0$ 	& $0$ 	& $0.44$& $0$ 	& $0$ 	& $0$ 	& $1.18$& $0$ 	& $0$	& $0$ 	& $0$ 	& $7.75$& $0.81$& $0$ 	& $0.02$\\
$( \rep{1}, \rep{2})_{\nicefrac{5}{6}}$ 	& $0$ 	& $0$ 	& $0$ 	& $0$ 	& $0$ 	& $0$ 	& $0$ 	& $0$ 	& $0$	& $0$ 	& $0$ 	& $0.00$& $0$ 	& $0$ 	& $0$ 	\\
$( \rep{1}, \rep{2})_{\nicefrac{1}{14}}$ 	& $0$ 	& $0$ 	& $0$ 	& $5.00$& $0$ 	& $0$ 	& $0$ 	& $0$ 	& $0$	& $0$ 	& $0$ 	& $0$ 	& $0$ 	& $0$ 	& $0$ 	\\
$( \rep{1}, \rep{2})_{\nicefrac{5}{14}}$ 	& $0$ 	& $0$ 	& $0$ 	& $1.00$& $0$ 	& $0$ 	& $0$ 	& $0$ 	& $0$	& $0$ 	& $0$ 	& $0$ 	& $0$ 	& $0$ 	& $0$ 	\\
\hline
$( \rep{1}, \rep{1})_{\nicefrac{1}{2}}$ 	&$18.79$&$23.81$&$14.60$& $0$ 	&$23.19$&$20.92$&$12.45$&$13.59$&$18.80$&$29.37$&$29.20$& $0$ 	&$25.33$&$37.34$&$41.21$\\
$( \rep{1}, \rep{1})_{\nicefrac{1}{3}}$ 	& $0$ 	& $0$ 	& $3.14$& $0$ 	& $0$ 	& $0$ 	& $7.72$& $0$ 	& $0$   & $0$ 	& $0$ 	&$42.99$& $5.22$& $0$ 	& $0.23$\\
$( \rep{1}, \rep{1})_{\nicefrac{2}{3}}$ 	& $0$ 	& $0$ 	& $0.43$& $0$ 	& $0$ 	& $0$ 	& $1.25$& $0$ 	& $0$	& $0$ 	& $0$ 	& $8.30$& $0.84$& $0$ 	& $0.03$\\
$( \rep{1}, \rep{1})_{\nicefrac{1}{4}}$ 	&$14.90$& $0$ 	& $0$ 	& $0$ 	& $0.07$& $0.17$& $0.68$& $0$ 	& $0$	& $0.03$& $0$ 	& $0$ 	& $0$ 	& $0.10$& $0$ 	\\
$( \rep{1}, \rep{1})_{\nicefrac{3}{4}}$ 	& $1.18$& $0$ 	& $0$ 	& $0$ 	& $0.00$& $0.01$& $0.04$& $0$ 	& $0$	& $0.00$& $0$ 	& $0$ 	& $0$ 	& $0.01$& $0$ 	\\
$( \rep{1}, \rep{1})_{\nicefrac{1}{6}}$ 	& $0$ 	& $0$ 	& $0.13$& $0$ 	& $0$ 	& $0$ 	& $0.63$& $0$ 	& $0$	& $0$ 	& $0$ 	& $0$ 	& $0.17$& $0$ 	& $0$ 	\\
$( \rep{1}, \rep{1})_{\nicefrac{5}{6}}$ 	& $0$ 	& $0$ 	& $0$ 	& $0$ 	& $0$ 	& $0$ 	& $0.01$& $0$ 	& $0$	& $0$ 	& $0$ 	& $0$ 	& $0.00$& $0$ 	& $0$ 	\\
$( \rep{1}, \rep{1})_{\nicefrac{1}{7}}$ 	& $0$ 	& $0$ 	& $0$ 	&$22.00$& $0$ 	& $0$ 	& $0$ 	& $0$ 	& $0$	& $0$ 	& $0$ 	& $0$ 	& $0$ 	& $0$ 	& $0$ 	\\
$( \rep{1}, \rep{1})_{\nicefrac{2}{7}}$ 	& $0$ 	& $0$ 	& $0$ 	&$16.00$& $0$ 	& $0$ 	& $0$ 	& $0$ 	& $0$	& $0$ 	& $0$ 	& $0$ 	& $0$ 	& $0$ 	& $0$ 	\\
$( \rep{1}, \rep{1})_{\nicefrac{3}{7}}$ 	& $0$ 	& $0$ 	& $0$ 	&$11.00$& $0$ 	& $0$ 	& $0$ 	& $0$ 	& $0$	& $0$ 	& $0$ 	& $0$ 	& $0$ 	& $0$ 	& $0$ 	\\
$( \rep{1}, \rep{1})_{\nicefrac{4}{7}}$ 	& $0$ 	& $0$ 	& $0$ 	& $9.00$& $0$ 	& $0$ 	& $0$ 	& $0$ 	& $0$	& $0$ 	& $0$ 	& $0$ 	& $0$ 	& $0$ 	& $0$ 	\\
\hline
\end{tabular}
}}%
\caption{Average numbers of vector-like exotics for MSSM-like orbifold models. 
Hypercharge is normalized such that $( \rep{3}, \rep{2})_{\nicefrac{1}{6}}$ is a left-chiral 
quark-doublet. The row ``\# MSSM'' lists the number of inequivalent MSSM-like orbifold models in 
our dataset. A complex representation has to be amended by its complex conjugate, e.g.\ 
$( \rep{3}, \rep{2})_{\nicefrac{1}{6}}$ stands for $( \rep{3}, \rep{2})_{\nicefrac{1}{6}} \oplus (\crep{3}, \rep{2})_{\nicefrac{-1}{6}}$.}
\label{tab:MSSMsummary2}
\end{table}

\clearpage

\providecommand{\bysame}{\leavevmode\hbox to3em{\hrulefill}\thinspace}

\end{document}